# How Secondary School Girls Perceive Computational Thinking Practices through Collaborative Programming with the Micro:bit

Mojtaba Shahin[1][*], Christabel Gonsalvez[1], Jon Whittle[2], Chunyang Chen[1], Li Li[1], Xin Xia[1]
[1] Faculty of Information Technology, Monash University, Australia
[2] CSIRO's Data61, Clayton, Australia
mojtaba.shahin@monash.edu, chris.gonsalvez@monash.edu, jon.whittle@data61.csiro.au,
chunyang.chen@monash.edu, li.li@monash.edu, xin.xia@monash.edu

**ABSTRACT**

Computational Thinking (CT) has been investigated from different perspectives. This research aims to investigate how secondary school girls perceive CT practices -- the problem-solving practices that students apply while they are engaged in programming -- when using the micro:bit device in a collaborative setting. This study also explores the collaborative programming process of secondary school girls with the micro:bit device. We conducted mixed-methods research with 203 secondary school girls (in the state of Victoria, Australia) and 31 mentors attending a girls-only CT program (OzGirlsCT program). The girls were grouped into 52 teams and collaboratively developed computational solutions around realistic, important problems to them and their communities. We distributed two surveys (with 193 responses each) to the girls. Further, we surveyed the mentors (with 31 responses) who monitored the girls, and collected their observation reports on their teams. Our study indicates that the girls found "debugging" the most difficult type of CT practice to apply, while collaborative practices of CT were the easiest. We found that prior coding experience significantly reduced the difficulty level of only one CT practice - "debugging". Our study also identified six challenges the girls faced and six best practices they adopted when working on their computational solutions.

## 1 Introduction

Computational Thinking (CT) has been widely researched due to its benefits for public services (e.g., education and healthcare), business sectors (e.g., financial markets), and society generally [4, 5]. The current research on CT has mostly focused on integrating CT into academic disciplines (e.g., biology [6] and mathematics [7]), the K-12 curriculum (e.g., [6, 8]), and developing programming environments and tools to promote CT skills (e.g., [9, 10]). Some research also examines and measures the learning outcomes of learners in three dimensions of CT: computational concepts, computational practices, and computational perspectives [10-12]. Others investigated the (frequency of) learning barriers (e.g., programming syntax, debugging) that students encounter in programming courses in academic settings and coding clubs (e.g., [13, 14]). According to Lye and Koh [10], most of the reported research in this area has focused on assessing the learning outcomes in terms of computational concepts such as the works done by [15, 16], but there are very few studies examining and assessing the learners' ability to develop and apply computational practices [17], and computational perspectives [18]. Reviews [3, 10, 11] have emphasised that there is a need for more empirical research to examine computational practices and perspectives in K-12 settings. This can be mainly justified by the fact that computational practices and perspectives are bigger contributors than computational concepts in achieving the main goal of introducing

---

[*] Corresponding author at: Faculty of Information Technology, Monash University, Australia. Tel: +61 3 99059455. E-mail address: mojtaba.shahin@monash.edu (M. Shahin)

CT through programming in K-12 settings, providing students with a set of skills (e.g., problem-solving skill) that they can leverage in their daily lives [10, 19, 20]. Furthermore, the existence of different definitions of CT in the literature implies that little consensus exists regarding the nature and the relative importance of particular CT practices [21, 22]. For example, young learners may individually develop one CT practice (e.g., debugging) within a particular period of time [23], whilst other CT practices such as collaborative problem-solving require mutual engagement and develop gradually over time, which might also be influenced by broader sociocultural factors [24]. This also indicates that learners experience different levels of difficulty when developing and applying particular CT practices [23, 25]. Despite the ongoing call to research CT practices, there have been few rigorous empirical investigations into CT practices and even less (if any) into the difficulty level of CT practices from the learners' perspective in K-12 settings [11], especially from the viewpoint of girls.

The importance of CT practices in daily life and a wide range of disciplines and professions has motivated educational researchers, governments, and practitioners to attempt to achieve "CT for all" goal [26, 27]. A growing number of interventions in the forms of CT programs, coding clubs, or computer science classes have been designed and delivered to move toward full participation in a computational world. Although such interventions target both men and women, lower representation by women continues to be an issue in the educational and professional worlds of computing and STEM fields [9, 28, 29]. While an important body of literature (e.g., [25, 30, 31]) shows that there is overwhelmingly more similarity than difference between girls and boys in terms of skills, competence, and achievement in computing, several factors deter girls from pursuing, choosing, and persisting in computing education and career paths [32]. These factors are varied but can be generally categorised into *psychological factors* (e.g., gender stereotypes such as "girls lack computing skills" or "people in the computing arena are geeky"), *social factors* (e.g., the influence of parents and peers and a lack of female role models), and *structural factors* (e.g., exposing girls to computing environments and curriculums and learning pedagogies that are uncomfortable for them) [32-35]. The above factors can significantly affect girls' perceptions, confidence, and interests in computing [32, 34]. A promising approach to attempt to address this challenge is girls-only after school interventions [26, 32, 36-38]. In contrast to mixed-gender computing education programs, girls-only education programs provide safer and more comfortable environments for girls to develop intention and a positive view towards computing and boost their confidence in computing [32, 33, 35, 39]. The question of *when* and *how* to present computational opportunities to women to increase participation is of interest. Research [28, 40, 41] suggests that the optimal time for cultivating girls' interest in STEM and computing is during late childhood and early adolescence, as they have more authority to take the opportunities they are interested in [42]. Furthermore, it is argued that exposure of women to developing computational ideas and solutions to realistic problems in social and collaborative contexts can be crucial motivating factors for them to pursue a STEM and/or computing education or career in the future [32, 37, 43, 44].

This work aims to understand how CT practices are perceived by secondary school girls (14-16 years old) when developing and implementing computational ideas and solutions with the micro:bit device[1], in a problem-based learning context in a collaborative setting [45]. More specifically, our research (a) investigates how secondary school girls perceive the difficulty level of CT practices including planning, decomposition, abstraction, generalization, algorithm, testing and debugging, and collaboration, which are more likely to emerge and be evaluated while students engage in programming activities; (b) investigates the impact of secondary school girls' prior coding experience on the perceived difficulty level of CT practices; (c) describes the challenges and barriers that secondary school girls experience when developing and implementing computational ideas and solutions; and (d) identifies the practices and techniques that secondary school girls learn, develop, and apply to overcome these challenges and barriers.

To that end, we propose the following research questions:

**RQ1.** What are the perceptions of secondary school girls on the difficulty level of CT practices when doing collaborative programming with the micro:bit?

**RQ2**. Is there a relationship between secondary school girls' prior coding experience and their perceptions of the difficulty of CT practices?

**RQ3.** What challenges do secondary school girls face when collaboratively implementing computational ideas with the micro:bit?

---
[1]https://microbit.org/

**RQ4.** What practices do secondary school girls employ to overcome these challenges?

We performed mixed-methods research, which collected data from the participants of a girls-only CT program (i.e., it is referred to as the OzGirlsCT program in this paper) to answer our research questions. a) We distributed two surveys (with 193 valid responses each) to 203 secondary school girls who participated in the OzGirlsCT program. b) We conducted another survey with 31 valid responses from 31 mentors who guided and closely monitored the girls during the OzGirlsCT program. 3) We asked the mentors to provide their observations (i.e., in total, 52 observation reports) on the work habits, behaviours, and experiences of the girls in their teams during the OzGirlsCT program.

The main findings of this study are: (1) our participants (i.e., secondary school girls and mentors) quantitatively indicate that "*debugging*" is the most difficult type of CT practice to apply, followed by "abstraction". (2) Collaborative practices of CT are the easiest practices to apply. (3) Prior knowledge and experience of coding can significantly reduce the difficulty level of "*debugging*"; (4) The challenges that the secondary school girls face when developing and implementing computational solutions with the micro:bit can be attributed to "*incorporating idea into the micro:bit*", "*code debugging*", "*code complexity*", "*the micro:bit limitations*", "*personality traits*", and "*coding experience*"; and (5) the main practices employed by the girls to overcome the challenges are "*feedback-driven development*", "*establishing a collaborative and supportive culture within the team*", "*simple design, better code*", "*predictive thinking*", "*prioritising quantity over quality*", and "*leveraging external resources*".

The key contributions of this study are summarised as follows:

- A relatively large-scale study that employs both quantitative and qualitative analyses to understand how secondary school girls perceive CT practices through collaborative programming with the micro:bit;
- A better understanding of the difficulty level of CT practices;
- An empirical investigation into the collaborative programming process of secondary school girls in their early efforts in programming;
- Concrete and actionable implications for educational researchers, practitioners, and policymakers.

This paper is organised as follows: In Section 2, we describe the background and related work. Section 3 details our research method, and our findings are reported in Section 4. Our discussion and reflection on the findings are presented in Section 5. Finally, Section 6 summarises our study.

## 2 Background and Related Work

This section reports some background for the research presented in this study, along with a brief discussion of the related studies.

### 2.1 Computational Thinking Definition and Scope

Computational Thinking (CT) is increasingly acknowledged as a set of fundamental skills to nurture, equip, and inspire the next generation for the workforce of the digital era [46-48]. The idea behind CT was introduced by Seymour Papert [49] and then popularized by Jeannette Wing [20]. Despite being researched for almost two decades, there is no single, overarching definition of CT, and little consensus exists as to what skills and competencies constitute CT [3, 12, 50]. Whilst Aho [51] conceptualises CT as "*the thought processes involved in formulating problems so their solution can be represented as computational steps and algorithms*", CT is viewed by Cuny et al. [50] as a skill set which everyone should develop to formulate and solve problems like a computer scientist. The promising benefits of CT have stimulated widespread interest among educational researchers, practitioners, and policymakers to integrate and implement CT into STEM (Science, Technology, Engineering, and Mathematics) curricula and K-12 education [10, 52, 53]. CT is referenced as a core component of STEM disciplines, in particular, the Computer Science (CS) discipline [9, 26]. In addition to the efforts to define CT, Brennan and Resnick [12] developed a framework for operationalising CT in K-12 education. The framework has three dimensions: computational concepts (i.e., the concepts such as *sequences* and *conditional statements* that learners use in their program), computational practices (i.e., the problem-solving practices such as *decomposition* and *debugging* that learners develop and apply while they are engaged in programming), and computational perspectives (i.e., this dimension includes the perspectives that are formed by learners about themselves and the world around them).

## 2.2 Computational Thinking Practices and Computing Teaching

As noted in Section 2.1, there is no agreement on what practices matter in CT. Some researchers use different terminology, such as CT skills or CT competencies, to refer to CT practices. Brennan and Resnick [12] suggest that the four core practices in CT are "abstracting and modularizing", "reusing and remixing", "testing and debugging", and "being incremental and iterative". Grover and Pea [9] refer to a broader list of skills than of Brennan and Resnick [12] as CT skills: "abstractions and pattern generalizations", "systematic processing of information", "symbol systems and representations", "algorithmic notions of flow of control", "structured problem decomposition (modularizing)", "iterative, recursive, and parallel thinking", "conditional logic", "efficiency and performance constraints", and "debugging and systematic error detection". On the other hand, Korkmaz et al. [54] consider creativity, critical thinking, and cooperativity as CT practices. Still, others refer to communication and working effectively in teams as key practices in CT [16, 55, 56].

Researchers have investigated the process of developing and acquiring CT skills from different perspectives such as age, gender, and pedagogical strategies. Atmatzidou and Demetriadis [25] explored how generalization, algorithms, abstraction, decomposition, and modularity skills are developed in the context of robotics among students grouped based on age and gender. Whilst the study found that age and gender did not impact the development level of CT skills, it has shown that girls needed more time and effort to achieve the same skill level. Similarly, Durak and Saritepeci [30] indicated that the gender of students did not affect their CT skill levels. Doleck et al. [57] empirically showed that there was no significant association between academic performance and four CT skills, including creativity, critical thinking, algorithmic thinking, and problem-solving, but their study only found that cooperativity as a CT skill had a negative relationship with academic performance. In another study [30], it was found that these four CT skills could be highly predicted by academic success in mathematics classes. From the teacher's perspective, Günbatar [58] evaluated the same set of CT skills and found that in-service teachers were significantly better in the development of this set of CT skills compared to pre-service teachers, except for problem-solving. Lewis [59] conducted a case study to understand the behaviour of sixth-grade students in the debugging process. He observed that a key competence in debugging is identifying and paying attention to the important elements of code state. He also found that having domain knowledge and such competence mediates the debugging process.

Aivaloglou and Hermans [14] studied teachers' perspectives about code clubs and revealed that debugging and abstract thinking were the most frequent programming learning barriers for students attending code clubs. The study [14] also found that girls were better than boys in collaboration and communication skills. In another study [13], Dorn et al. collected data about learning barriers in programming from teachers and computer science first-year students before the first lecture started. It was found that "way of thinking" (e.g., abstract thinking, complex thinking, logical thinking) was the most common barrier perceived by the students. At the same time, the teachers believed that "programming language/syntax" and "diligence/commitment/stubbornness" were the most common challenges. However, both groups indicated that debugging was difficult but less frequently mentioned by both groups.

Some studies (e.g., [60-62]) indicated the positive impact of prior coding/computing experience on students' performance in computer science courses. Alvarado et al. [60] observed this positive impact on student grades in introductory and advanced courses in computer science, revealing students with pre-college computing experience performed significantly better than their peers with less or without experience in these courses. Wilcox and Lionelle [61] achieved the same findings for the introductory computer science course but found that the impact of computing experience in advanced courses was gradually diminished. On the other hand, Durak and Saritepeci [30] found that the experience of secondary and high school students using ICT did not impact their CT skill levels.

Although CT skills are usually introduced and evaluated through computer programming activities in schools, several studies have claimed that the unplugged approach (i.e., when digital devices are not used) is an effective approach for this purpose [17, 63, 64]. By investigating pattern recognition, algorithmic design, decomposition, and abstraction, Brackmann et al. [63] showed that learners who participated in the unplugged activities developed this set of CT practices significantly more than those who were not engaged with these unplugged activities.

Our work is different from the existing studies: (1) our findings come from surveying 193 secondary school girls and a survey and an observation report completed by 31 mentors rather than only *school students* [59], *teachers* [14], or *university students* [13, 60, 61]. This study [13] collected data from both students and teachers; however, it targeted first-year university students, not secondary school students. (2) The studies [13, 14, 59] either focused on *one CT practice* or *a few particular CT practices* (e.g., debugging, abstract thinking) and indicated that how frequent they are reported as a learning barrier in programming among many other types of learning barriers (e.g., programming syntax). Our study, however, examines the difficulty level of 12 CT practices (**RQ1**). (3) The studies [60, 61] focused on the impact of prior coding/computing experience on *student grades* in computer science courses. In contrast, our work assesses the impact of coding experience on the difficulty level of 12 CT practices (**RQ2**). (4) Finally, our study provides a rigorous exploration and analysis of the programming process of secondary school girls with the micro:bit device from the perspective of the challenges that they may face in their early programming efforts and the best practices that they develop and apply to overcome these challenges (**RQ3** and **RQ3**).

For the sake of consistency and readability, we adopted the term *CT practice*, as suggested by Brennan and Resnick [12], instead of *CT skill* or *CT competence* in this study.

## 2.3 Problem-based Learning

Recently, researchers and practitioners have shown more interest in designing learning environments that provide young learners an opportunity to learn and develop CT concepts and practices through working on problems that are *authentic* and *relevant to their lives* and *their interests* [10, 43]. This consideration is also stressed by researchers from other disciplines [65, 66]. Tissenbaum et al. [43] referred to this trend as *computational action*, which can lead to more intellectual engagement of the learner in problem-solving activities [67]. Whilst Hsu et al. [68] revealed that CT can fit into many learning strategies such as "problem-based learning", "project-based learning", and "game-based learning", we found a strong synergy between problem-based learning (PBL) and this new trend in CT, as PBL deals with ill-structured, realistic problems that have a significant impact on learners' lives, and chose to use this learning strategy for the OzGirlsCT program.

PBL is a type of experiential learning in which the learning process of students is facilitated by teachers [45]. Given the increasing importance of the ability to both identify problems and provide solutions to solve the problems in the 21$^{st}$ century, PBL as an instructional approach situates learning in real-world problems and offers the potential to help learners identify real-life, ill-structured problems, develop solutions, and construct knowledge [69]. Learners work in collaborative groups and take responsibility for their learning in the PBL approach. During the learning process, learners are responsible for identifying the learning issues (i.e., their insufficiencies and strengths in resolving a realistic, ill-structured problem). They have to discover what new knowledge they need to acquire to fix the problem (i.e., known as "self-directed learning"). This necessitates an extensive reflection on the knowledge being constructed and the effectiveness of the solutions proposed and employed [70]. In PBL, teachers (also known as tutors) only facilitate and activate the collaborative learning process and do not provide the information related to the problem [45, 69]. Hmelo-Silver [45] asserts that PBL can promote the construction of flexible knowledge and the development of skills such as problem-solving, collaboration, critical thinking, and self-directed.

## 2.4 Micro:bit

Programming environments and languages for education undergo tremendous change, and every year new environments and tools emerge. This compels educational researchers to investigate if and how the emerging learning environments and tools affect the learning outcomes of learners and the difficulties that they may experience when developing certain CT skills [3, 71]. The micro:bit is a "RAM-based programmable Internet of Things (IoT) device" created by the BBC to train children about programming and preliminary computing principles [72, 73]. The micro:bit was released formally in 2016 and has been used by more than 20 million children in 60 countries [74]. It is a low-cost, pocket-size device (4×5 cm) with an ARM Cortex-M0 Processor to execute the programs. The micro:bit device includes various features such as a 256KB flash memory, an accelerometer, 16KB RAM, two programable buttons, one reset button, and 25 programmable LEDs organised in a 5x5 grid (See Figure 1) [74]. There are three options to write a program with the micro:bit: Blocks, JavaScript, and Python. Users can create programs using a block-based programming language, dragging and dropping blocks in a logical order. Microsoft MakeCode[2] and MicroPython[3] are two official text-based code editors of the micro:bit. The former provides users with JavaScript programming experience, and the latter supports Python. These features characterise the micro:bit as a hybrid block/text programming environment [71], in which users can switch back-and-forth between the block-based and text-based editors. A wide range of educational projects can be implemented using the micro:bit device, ranging from developing classical and interactive games (e.g., rock, paper, scissors game), to building prototypes to track, monitor, and simulate objects in environments (e.g., a prototype to monitor how climate change affects animals), to helping people with special needs (e.g., a prototype tool to support autistic people in communicating with others) [74, 75].

---
[2]https://makecode.microbit.org/
[3]https://python.microbit.org/

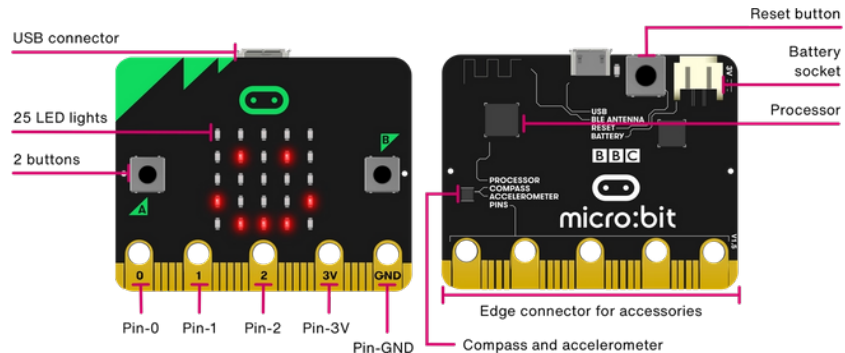

Figure 1. The front and back of the micro:bit device (taken from [74])

Most of the previous investigations of the micro:bit device are focused on fostering learners' enthusiasm and interest in computing and programming [73, 76], but no research has explored how learners develop and apply CT practices with the micro:bit as a hybrid block/text programming environment. In this study, we leveraged the micro:bit device and Microsoft MakeCode as our introductory programming environment for two reasons: 1) Whilst the micro:bit provides learners with hands-on experience of coding similar to dominating programming environments around CT (e.g., Scratch[4]), it also engages novice users with ubiquitous computing through providing insights into embedded systems and enabling them to understand how hardware and sensor-based devices work [77]. 2) In contrast to the existing devices such as Arduino[5] and Raspberry Pi[6], the micro:bit is specifically created for educational purposes. The micro:bit is intentionally designed to have fewer difficulties for novice programmers to learn and construct embedded systems (e.g., programmers do not require to run a full operating system) [77].

## 3 Methodology

The OzGirlsCT program was the first step of the "Women in STEM and Entrepreneurship" (WISE) program. The WISE program was a three-step education program with the following objectives: (1) understanding how girls develop and perceive CT; (2) developing a technology-focused entrepreneurial intention amongst girls [78]; and (3) increasing awareness of girls in STEM. The WISE program participants were Year 10 secondary school girls in the state of Victoria in Australia in 2019. Their ages ranged from 14 to 16. We refer to them as "girls" or "students" in this paper. This study only focuses on the OzGirlsCT program. In Australia, secondary schools last six years, and students should attend until age 17. Year 10 is the beginning of senior secondary school.

### 3.1 OzGirlsCT Program

The OzGirlsCT program involved three one-day workshops. We organised these workshops in three days in 2019. In total, 203 girls from 44 secondary schools, grouped into 52 teams of 3-4 students, participated in three workshops (i.e., each team attended only one workshop). Each team was mentored by a "big sister" university mentor (hereafter "mentor").

#### 3.1.1 OzGirlsCT Program recruitment

#### 3.1.1.1 Secondary School Girls

Information about the free OzGirlsCT program was sent to Victorian schools, and the schools advertised the OzGirlsCT program to their students through their internal communication channels. The schools used a range of methods to recruit participants within their schools. In some schools, teachers nominated students, while in others, students were able to self-nominate to the OzGirlsCT program. The students in each team were from the same school, and the formation of the teams was finalised by the schools. The friendship level of the students in the teams varied considerably, from those who did not know each other at all, to some of the team members in a team having close friendships. Initially, each school was able to send just one team; however, as we had additional capacity after the initial recruitment, schools that had requested participation from more than one team were able to send a second team.

---
[4]https://scratch.mit.edu
[5]https://www.arduino.cc
[6]https://www.raspberrypi.org

### 3.1.1.2 Mentors

Specifically, we sought women students at Monash University to recruit mentors for the OzGirlsCT program. We first designed an online Expression of Interest (EOI) form and advertised the EOI form through Monash University's newsletter, as well as sent it to the mailing list of IT women students at Monash University. In the EOI preamble, we described the purpose of our study and the characteristics we were looking for in potential mentors. Specifically, we sought women students who obtained or were doing a STEM-related degree at Monash University. They also had to have a basic level of programming skills. 66 women students completed the EOI. The next step involved interviewing the 66 candidates. 14 out of the 66 candidates did not attend the interviews. During the interviews, we asked the interviewees to describe the qualities and characteristics that make a good mentor for young girls. Then, we sought demographic information about the interviewees, including programming experience and mentoring, tutoring, or volunteering experience. We selected 31 women students as mentors. The main criteria used to select the 31 mentors were communication skills, personal characteristics (e.g., friendly and confident), tutoring and volunteering experience, and programming experience. Subsequently, we arranged a one-day training workshop to introduce the mentors to their responsibilities during the OzGirlsCT program and describe the OzGirlsCT program's objectives. Furthermore, mentors were trained on how to work and program with the micro:bit. Each mentor guided only one team per OzGirlsCT workshop. However, a few mentors guided teams across multiple workshops.

### 3.1.2 Pre-OzGirlsCT Program Activities

The only pre-requisite for entry into the OzGirlsCT program was the student's year level at school. We did not expect the students to have prior programming experience. A month before OzGirlsCT program workshops started, we provided introductory pre-work for the micro:bit to attempt to bring all students to the same level. More specifically, each student was given a micro:bit and encouraged to read the eBook created by our research team and do the exercises in the eBook. They were given resource links so that they could explore the micro:bit website [74] and a range of other tutorials and websites to learn about the micro:bit device features and how to code with the micro:bit. The eBook included several sample projects, tutorial videos, and exercises. Further, the student teams were also introduced to their mentors at this time and were encouraged to communicate with their mentors and ask any questions (e.g., debugging) regarding programming with the micro:bit, as they completed the pre-work. While we did not formally check that the activities were completed, students were told that they must complete all the required activities before their workshop, and were supported by their mentors to do so.

### 3.1.3 OzGirlsCT Program Activities

As described in Section 3.1.2, each student was given a micro:bit before the OzGirlsCT program started, and each student brought their micro:bit to the workshop. Students were also requested to bring their own laptops. Each workshop started with familiarizing students with the concepts of IoT. Next, students were asked to brainstorm the problems that they had in and/or observed around their everyday life. Following the suggestion proposed by Tissenbaum et al. [43], students were not given a predefined problem or task (e.g., design a game based on predefined specifications), but were encouraged to develop computational ideas and solutions around realistic problems that were important to them and their communities. The members of each team then worked together to decide on which problem to focus on. While we encouraged team members to collaborate with each other, teams were responsible for choosing their working style and collaboration strategy as we followed problem-based learning. The difficulty levels of the problems chosen by the girls widely varied. For example, one team decided to improve the safety of cyclists at night, and another team was concerned about distracted students in class. The next step included completing the following problem statement for the identified problem: "*We believe that [the identified problem] … is a problem for [who] … because [reason] …*". Then, teams had to propose feasible solutions to solve their identified problem. The proposed solutions needed to have an IoT component. After choosing a solution to the identified problem, teams were requested to complete the statement: "*We could solve this problem by [solution idea] … and *could demonstrate this on the micro:bit by [prototype idea] ...*". In the next step, teams utilised the micro:bit device to prototype the proposed ideas. We encouraged the girls to verify their ideas, statements, and solutions in all steps by seeking feedback from other teams. Finally, each team had to present a business pitch discussing their products.

## 3.2 Data Collection

To answer the research questions introduced in the Introduction section, we conducted a mixed-methods empirical study with a *concurrent triangulation strategy*, characterised by employing different data collection methods concurrently to confirm, cross-validate, and augment findings [79]. We collected data from the participants of the OzGirlsCT program using three surveys and observations. Figure 2 shows an overview of our research method. In the first step, a survey protocol was developed from the literature, grey literature, and the practical programming experience of the research team to collect data from students. Unlike formal literature, such as journal articles, grey literature refers to a body of materials such as government reports that have not been published and/or controlled by commercial publishers [80]. In addition, a survey protocol and an observation protocol were developed to collect data from mentors. In the next step, we ran two surveys to collect the students' perspectives on CT from different perspectives. Mentors were also asked to complete a survey and submit their observation reports.

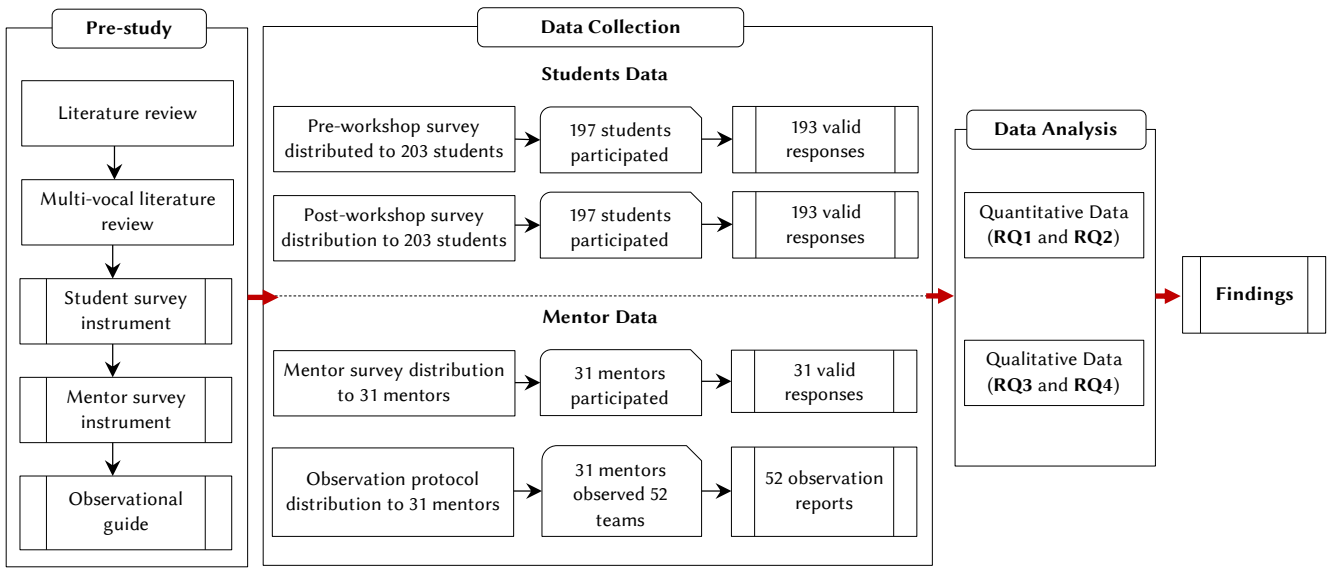

Figure 2. Research Method Overview.

### 3.2.1 Pre- and Post-workshop Student Surveys

**Protocol**: We designed two surveys to collect data around the background, challenges, behavioural patterns, practices, and experiences of girls during the OzGirlsCT program. We carried out the first survey (i.e., pre-workshop survey) at the beginning of the OzGirlsCT workshops. In total, the pre-workshop survey consisted of 33 questions. In this paper, we only used three questions of the pre-workshop survey, which are demographic questions around prior coding experience, type of school, and class performance (e.g., "*Do you have any prior computer coding-related experience? If so, how long?*"). The rest of the pre-workshop survey questions sought the students' skills and interest in entrepreneurship and STEM. At the end of each OzGirlsCT workshop, students were asked to fill out a post-workshop survey with 40 questions. All questions, except one, were mandatory. Following the high-level objectives of the WISE program, these 40 questions focused on entrepreneurship (e.g., entrepreneurial inspiration), STEM, and CT. The following list details the questions of the post-workshop survey which were used in this study:

- **CT practices**. This study focused on a set of CT practices, including *planning*, *decomposition*, *abstraction*, *generalization*, *algorithm*, *testing and debugging*, and *collaboration*. These practices are more likely to emerge and be evaluated when CT is introduced through programming activities [18, 22, 81]. We designed 12 items to collect students' views on these CT practices, using age-appropriate terminology to indirectly measure these practices.

Table 1 presents these CT practices, their respective items, and sources. We asked students to rate the difficulty of the 12 items for them (i.e., from *very difficult* to *very easy*).

- **Challenges and practices**. Through two mandatory open-ended questions, we asked students to share any challenges that they faced when implementing their ideas with the micro:bit and the practices or techniques that they used to overcome these challenges.
- **General comments**. Through an optional open-ended question, we let students share any general comments about the OzGirlsCT program.

**Participants**: The participants in the pre- and post-workshop surveys were the same Year 10 girls who attended the OzGirlsCT program workshops. Overall, 203 students participated in the workshops. Students came from different types of schools - public, private and Catholic, and most had good academic records as judged by their self-reporting. Whilst completing the surveys was voluntary, 197 out of 203 students completed each survey. We removed 4 responses from each survey because of the invalid nature of the responses (e.g., when we were not able to match a student's responses in the post-workshop survey to the pre-workshop survey due to Survey ID entry errors) [82]. In the end, we received 193 valid responses for each survey (See Figure 2).

Table 1. CT practices, their respective items (statements), and sources

| CT Practices | Item(s) | Sources |
| --- | --- | --- |
| Planning | CTP1. Planning an idea before implementing it with the micro:bit | [11, 83] |
| Decomposition | CTP2. Breaking down an initial idea into smaller, more manageable steps/parts | [25] |
| Abstraction | CTP3. Leaving out the irrelevant detail/information in the description of an idea | [25] |
| Generalization | CTP4. Developing a general solution that can be applied to other problems in the future | [25] |
| Algorithm | CTP5. Creating a series of ordered steps to implement an idea with the micro:bit | [25] |
| | CTP6. Exploring diverse solutions to an idea, until the ideal solution is achieved | |
| Testing and Debugging | CTP7. Testing code frequently to check if it works | [10, 11] |
| | CTP8. Identifying errors in code | |
| | CTP9. Finding a solution to fix the identified errors in code | |
| Collaboration | CTP10. Giving feedback to teammates and making suggestions to improve idea/code | [16, 55, 56] |
| | CTP11. Working collaboratively with team members | |
| | CTP12. Reaching a consensus in group decisions | |

### 3.2.2 Mentor Survey

**Protocol**: The data collected from the post-workshop survey is based on students' self-reporting and self-assessment, which might be unreliable [84]. To alleviate this limitation, we deployed an online survey with a similar goal of that of the post-workshop student survey to collect mentors' perspectives on the challenges, practices, and experiences of students during the OzGirlsCT program. For this study, mentors were asked to indicate the difficulty level of 12 CT practices for the team(s) that they mentored. These 12 CT practices were exactly the ones that were asked in the post-workshop student survey (See Table 1). Next, mentors shared the challenges that students faced when developing and implementing their ideas with the micro:bit and the practices used by students to address those challenges. We closed the mentor survey by asking mentors to share any comments and feedback they may have about the activities conducted during the OzGirlsCT workshops.

**Participants**: The participants in the mentor survey were the mentors recruited for the OzGirlsCT program (See Section 3.1.1.2). Mentors were asked to respond to the mentor survey at the end of the final OzGirlsCT workshop they participated in, as they may have participated in multiple workshops. We received 31 valid responses from the mentors.

### 3.2.3 Mentor Observations

**Protocol**: Although a great deal of data can be gathered through surveys, the data gathered from surveys may be subjective and include potential biases such as social desirability [85]. Hence, we used observation to collect actual and firsthand accounts about the work habits, behaviours, and interactions of students when developing and implementing their ideas with the micro:bit. Moreover, we wanted to identify the challenges and practices which students were unaware of or unable to communicate through the pre- and post-workshop surveys [85]. Considering the numbers - 203 students grouped into 52 teams, it was not possible for us, as the authors of this paper (the researchers), to conduct a participant-observation study [86]. Therefore, we asked mentors to act as *observers* as well. During the one-day training workshop for mentors (discussed in Section 3.1.1.2), we instructed mentors about observation techniques. From the data collection perspective, it was a semi-structured observation [85]. Whilst we provided an observation protocol to guide mentors on what to observe and collect, they were free to collect any data that they perceived as important. It is worth noting that mentors were not directly involved in their team's problem-solving process - their role was to guide and facilitate. The observation was accomplished with the think-aloud technique as students were asked to think out loud (i.e., verbalize their thought process) while working on their ideas [79]. Since students may have sometimes forgotten to verbalize, mentors, as the observers, reminded them occasionally (e.g., every 15 minutes) to continue thinking out loud. The observation happened at the team level. Mentors submitted their team observation at the end of each OzGirlsCT workshop. It should be noted that as some mentors guided more than one team (each team attending a different OzGirlsCT workshop), the number of observation reports correlated to the number of teams. In the end, we collected 52 observation reports from mentors. The following questions were central to the observation protocol:

- Describe your observations on the team's motivations, thoughts, and assumptions.
- Describe your observations on how the members of the team interacted, communicated, and collaborated.
- Describe your observations on the team's work habits.
- Describe your observations on the issues that the team faced.
- Describe your observations on how the team fixed the issues.

Besides the above questions, the observation protocol included three single-choice questions. We asked mentors to indicate which of the following statements best applied to the workstyle of the team they mentored:

- All students were contributing evenly.
- A dominant student was guiding work, other students contributing.
- A dominant student was influencing work contributions, other students contributing unevenly.
- A dominant student was determining team effort, but some contribution from other students.
- Low student contribution, relying on one 'leader' to carry work.

The mentors were also asked to select one of the statements below to show the communication patterns within their mentored teams in two timeslots. The first timeslot was *ideation* time when students worked on developing their ideas. The second timeslot included *coding* time, in which students implemented their ideas with the micro:bit.

- All students were communicating freely: critical evaluations, objections, critiques, and opinions were freely exchanged within the team.
- All students were communicating well: some critical evaluations, objections, critiques, and opinions were exchanged within the team.
- All students were communicating intermittently: critical evaluations, objections, critiques, and opinions were exchanged but only sporadically.
- All students were communicating inadequately: critical evaluations, objections, critiques, and opinions were voiced but only with prompting from the mentor.
- Little to no communication, ideas, and critiques were rationed by a dominant student.

## 3.3 Data Analysis

### 3.3.1 Quantitative analysis for RQ1 and RQ2

The close-ended questions, including the Likert scale and single-choice questions, were analysed using IBM SPSS Statistics 26 software to answer **RQ1** and **RQ2**. More specifically, we used the following statistical techniques: (i) We identified the most difficult CT practices for students (**RQ1**) by using the Scott-Knott Effect Size Difference (ESD) test proposed by Tantithamthavorn et al. [87]. We applied the Scott-Knott ESD test on the Likert scores of 12 CT practices from students' and mentors' perspectives. The main advantage of the Scott-Knott ESD test over the Scott-Knott test [88] is that it does not need normally distributed data. (ii) We applied the Mann-Whitney U test [89] to compare perceiving CT practices between students and mentors (**RQ1**). The Mann-Whitney U test was used because none of CT practices' scores was normally distributed (i.e., the Shapiro-Wilk test's p-values were less than 0.05 for all CT practices [90]). Also, the variables (i.e., CT practices) were measured at the ordinal level. Finally, a non-parametric Levene's test confirmed the equality of variances in both samples (i.e., p-values > 0.05 for all CT practices) [91, 92]. (iii) We conducted the Kruskal-Wallis one-way ANOVA tests [89] to understand the relationship between the difficulty level of CT practices and students' coding experience (**RQ2**). The homogeneity of variances in the studied samples was verified by the non-parametric Levene's test (i.e., p-values > 0.05 for all CT practices) [91, 92]. Furthermore, the pairwise post hoc tests (i.e., pairwise comparisons) [89, 93] were carried out on each pair of groups.

### 3.3.2 Qualitative Analysis for RQ 3 and RQ4

We analysed the answers to the open-ended questions using open coding and constant comparison as the two main qualitative data analysis techniques of Grounded Theory (GT). The collected qualitative data was used to answer **RQ3** and **RQ4**. The qualitative analysis was supported by the NVivo software[7] [94, 95]. GT enables the researcher to identify the main concerns of the participants and understand how they address the concerns. This is achieved by constantly comparing data while the levels of abstraction are increasing [94, 95].

We first created three top-level nodes in NVivo according to our data sources (See Figure 3): (1) student survey data, (2) mentor survey data, and (3) observation data. Since **RQ3** was about challenges and **RQ4** focused on practices, each high-level node (e.g., mentor survey data) was further decomposed into two sub-nodes: *challenge* node and *practice* node. Subsequently, the first author extracted the statements (e.g., **Statement_2** in Figure 3) about the challenges that students faced and the statements (e.g., **Statement_1** and **Statement_3** in Figure 3) about the best practices and techniques adopted by students in each of the data sources. Note that some statements (e.g., **Statement_1**) included both challenge(s) and the practice(s) used to address the challenge(s). Hence, we placed such statements in the *challenge* node and *practice* node to maintain the relationship between challenges and practices. In the next step, he performed open coding over multiple iterations to thoroughly analyse the data gathered from each data source. This step resulted in capturing key points in our data sources and assigning a label (i.e., code) to each key point. Figure 3 shows the process of applying open coding on **Statement_1** and **Statement_3** identified two codes for the *practice* node. The analysis of **Statement_1** and **Statement_2** led to adding two codes to the *challenge* node (See Figure 3).

| Statement_1 from Student Survey | Statement_2 from Mentor Survey | Statement_3 from Mentor Observation |
|---|---|---|
| **Raw data**: *"It was originally very hard to incorporate the usage of the micro: bit in our original design, however, after we changed our design a few times and came up with different ideas, we were able to use the micro: bit effectively"*.<br><br>**Key point**: *"Hard to the original design into our original design", "Changing the original design to the new ones and select the feasible one for micro:bit"*<br><br>**Code**: Incorporating idea into the micro:bit, Generating as many as possible ideas/solutions | **Raw data**: *"Too scared to explore functions and do things on their own"*.<br><br>**Key point**: *"Being scared"*<br><br>**Code**: Fear | **Raw data**: *"They tried to limit scope creep by forcing themselves to focus on the most important functionalities and then listed the extra functionalities for next steps...."*.<br><br>**Key point**: *"Focusing on the most important functionalities", "Extra functionalities for next steps"*<br><br>**Code**: Incremental approach |

Figure 3. Examples of constructing codes

The next step included performing the constant comparison technique to compare all codes identified in a data source against each other as well as to compare them with the codes from other data sources [96]. The identified codes from the previous step were iteratively grouped to generate *concepts*, and then the generated *concepts* were used to create *categories* [95]. In the next step, the identified *codes*, *concepts*, and *categories* were shared with the second author for review. Then, the first author and the second author held several face-to-face meetings to discuss the *codes*, *concepts*, and *categories* and solve any disagreements and inconsistencies and reach a consensus on the final list of *codes*, *concepts*, and *categories*. Figure 4 shows how performing the constant comparison technique on four concepts produced the category, "*Establishing a collaborative and supportive culture within the team*".

---

[7]http://www.qsrinternational.com

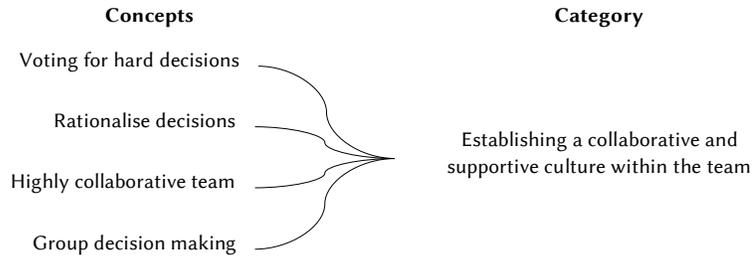

Figure 4. Building a category by applying constant comparison

## 4 Results

This section presents the findings of the RQs. It should be noted that when we refer to data from the student survey, we use **SSXY** notation. In our notation, **X** (1…52) shows the team number of a student, and **Y** (1…4) refers to the student number in a team. For instance, **SS153** refers to *student 3* within *team 15* from the student survey. The participants (mentors) in the mentor survey and observation reports are presented as **MSX** and **MOY**, respectively, in which **X** (1…31) is the participant number and **Y** (1…52) indicates the team number. For instance, an excerpt from the observation report of *team 37* is marked as **MO37**.

### 4.1 Secondary School Girls Demographic Data

Figure 5 shows an overview of students' demographics. 97 out of 193 (50.2%) students had at least one month of coding experience, 32 students (16.5%) had less than 1-month of coding experience, while 69 students (35.7%) had no coding experience before the OzGirlsCT program. 84 (43.5%) students came from Government schools. The rest from Catholic schools (54 girls, 27.9%) and Independent schools (55 girls, 28.4%). Almost 97% of the girls designated that their academic performance in the class ranged from 70%-79.99% to 90%-100%.

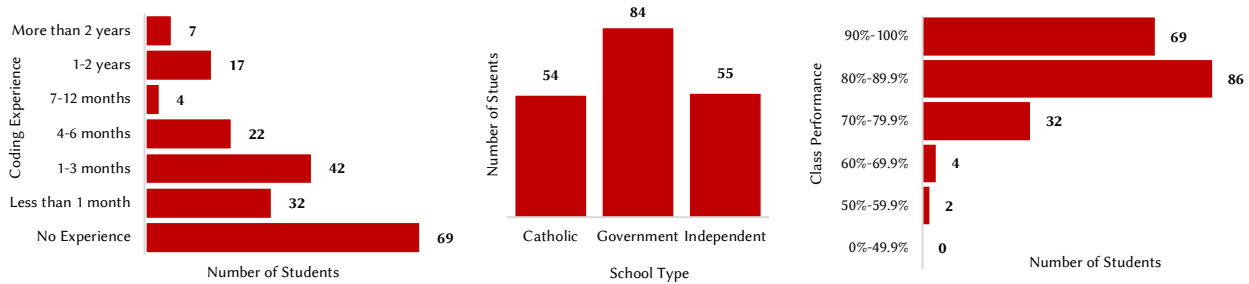

Figure 5. Secondary School Girls' demographic data

### 4.2 What are the perceptions of secondary school girls on the difficulty level of CT practices when doing collaborative programming with the micro:bit? (RQ1)

To understand our participants' perspectives on CT practices, we provided 12 statements (See Table 1) for which students could indicate the difficulty level of CT practices and mentors could indicate the difficulty level of CT practices for the team(s) that they mentored. Another goal of these statements is to understand whether the perceptions of CT practices differ based on the participant role (students vs. mentors). To identify the most difficult CT practices for students, we performed the Scott-Knott Effect Size Difference (ESD) test [87] on the Likert scores of 12 CT practices from students' and mentors' perspectives. Table 2 shows that CT practices are classified into 6 statistically different groups in terms of difficulty for all respondents. In Table 2, a group with a smaller number has more difficult CT practices. We observe that CTP9, CTP8, and CTP3 are among the most difficult practices (Group 1) for students to apply. Interestingly, 2 of these CT practices (CTP9 and CTP8) are related to testing and debugging. Table 3 presents the mean and median Likert scores of 12 CT practices. As shown in Table 3, students' average Likert scores of CTP8 ("*identifying errors in code*") and CTP9 ("*finding a solution to fix the identified errors in code*") are 3.18 and 3.15 respectively, while mentors scored CTP8 (average Likert score: 2.77) and CTP9 (average Likert score: 2.87) lower. We also find that CTP11 ("*working collaboratively with team members*"), CTP12 ("*reaching* *a consensus in group decisions*"), and CTP10 ("*giving feedback to teammates and making suggestions to improve idea/code*") are among the top 3 easiest CT practices to apply.

Table 2. CT practices are grouped into six statistically distinct groups in terms of difficulty (Scott-Knott Effect Size Difference test applied to all respondents) – the smaller the group number, the more difficulty the CT practices

| Group | CT Practices |
|---|---|
| 1 | **CTP9**: Finding a solution to fix the identified errors in code |
|   | **CTP8**: Identifying errors in code |
|   | **CTP3**: Leaving out the irrelevant detail/information in the description of an idea |
| 2 | **CTP5**: Creating a series of ordered steps to implement an idea with the micro:bit |
|   | **CTP6**: Exploring diverse solutions to an idea, until the ideal solution is achieved |
|   | **CTP4**: Developing a general solution that can be applied to other problems in the future |
|   | **CTP1**: Planning an idea before implementing it with the micro:bit |
|   | **CTP2**: Breaking down an initial idea into smaller, more manageable steps/parts |
| 3 | **CTP7**: Testing code frequently to check if it works |
| 4 | **CTP10**: Giving feedback to teammates and making suggestions to improve idea/code |
| 5 | **CTP12**: Reaching a consensus in group decisions |
| 6 | **CTP11**: Working collaboratively with team members |

Figure 6 shows how students and mentors rated the difficulty level of each of the CT practices. As a student, the majority of the respondents claimed that "*working collaboratively with team members*" (i.e., 87% rated CTP11 as *easy* or *very easy*) and "*reaching a consensus in group decisions*" (i.e., 80% rated CTP12 as *easy* or *very easy*) are the easiest CT practices to apply. Mentors had the same feeling, as at least 87% of mentors believed that applying CTP11 and CTP12 is an easy or very easy task for students. On the other hand, the following were rated by students as the most difficult CT practices to apply: (1) finding a solution to fix the identified errors in code (CTP9: 29% *difficult* or *very difficult*, 34% *neutral*, 37% *easy* or *very easy*); (2) identifying errors in code (CTP8: 28% *difficult* or *very difficult*, 34% *neutral*, 38% *easy* or *very easy*); and (3) leaving out the irrelevant detail/information in the description of an idea (CTP3: 24% *difficult* or *very difficult*, 38% *neutral*, 37% *easy* or *very easy*). Mentors had slightly different observations on the most challenging CT practices for students. They ranked CTP8 (42% *difficult* or *very difficult*, 23% *neutral*) as the most difficult CT practice for students to implement, followed by CTP1 (35% *difficult* or *very difficult*, 19% *neutral*) and CTP9 (32% *difficult* or *very difficult*, 26% *neutral*). In contrast to mentors, it is a commonly held belief among students that "*planning an idea before implementing it with the micro:bit*" (CTP1) is a relatively easy task, as only 16% of them rated this CT practice as *difficult* or *very difficult*. In addition, CPT6, CTP7, CPT10, CTP11, and CTP12 are the only CT practices that over 50% of students and mentors mutually believed are (very) easy practices to apply. Interestingly, these CT practices, except one (CTP7), can be classified as soft skills [97].

Not all practices were ranked as *very difficult*. Regarding which practices the respondents did not perceive as *very difficult* during ideation and coding, we have "*breaking down an initial idea into smaller, more manageable steps/parts*" (14% of students and 26% of mentors rated CPT2 as *difficult*) and "*reaching a consensus in group decisions*" (6% of students and 3% of mentors rated CPT12 as *difficult*). Furthermore, CPT10 ("*giving feedback to teammates and making suggestions to improve idea/code*") and CTP5 ("*creating a series of ordered steps to implement an idea with the micro:bit*") were never ranked as *very difficult* by students and mentors respectively.

The fifth column in Table 3 presents the mean difference in the average Likert scores of students and mentors. We were interested in understanding whether there is a significant difference in perceiving CT practices between students and mentors. Table 3 shows the results of the Mann-Whitney U test. We observe that there is a significant difference only for CTP2 (*U=2227.0, N1=193, N2=31, p-value=0.014, r=0.164*) [98], indicating that the difficulty level of "*breaking down an initial idea into smaller, more manageable steps/parts*" was perceived differently by mentors (median=3, mean rank=93.73) than by students (median=4, mean rank=115.52), with mentors seeing it as being significantly more difficult.

Table 3. CT practices and the mean and median Likert scores (very difficult= 1, difficult = 2, neutral = 3, easy =4, very easy = 5).

| ID | Computational Thinking Practices | Student (n=193) Mean (Median) | Mentor (n=31) Mean (Median) | Mean Difference | p-value |
|---|---|---|---|---|---|
| CTP1 | Planning an idea before implementing it with the micro:bit | 3.54 (4) | 3.16 (3) | 0.38 | 0.068 |
| CTP2 | Breaking down an initial idea into smaller, more manageable steps/parts | 3.57 (4) | 3.16 (3) | 0.41 | **0.014*** |
| CTP3 | Leaving out the irrelevant detail/information in the description of an idea | 3.19 (3) | 3.10 (3) | 0.09 | 0.624 |
| CTP4 | Developing a general solution that can be applied to other problems in the future | 3.53 (4) | 3.26 (4) | 0.27 | 0.234 |
| CTP5 | Creating a series of ordered steps to implement an idea with the micro:bit | 3.46 (3) | 3.10 (3) | 0.36 | 0.055 |
| CTP6 | Exploring diverse solutions to an idea, until the ideal solution is achieved | 3.40 (4) | 3.45 (4) | 0.05 | 0.707 |
| CTP7 | Testing code frequently to check if it works | 3.64 (4) | 3.68 (4) | 0.4 | 0.640 |
| CTP8 | Identifying errors in code | 3.18 (3) | 2.77 (3) | 0.41 | 0.115 |
| CTP9 | Finding a solution to fix the identified errors in code | 3.15 (3) | 2.87 (3) | 0.28 | 0.208 |
| CTP10 | Giving feedback to teammates and making suggestions to improve idea/code | 3.80 (4) | 3.90 (4) | 0.1 | 0.236 |
| CTP11 | Working collaboratively with team members | 4.34 (5) | 4.48 (5) | 0.14 | 0.144 |
| CTP12 | Reaching a consensus in group decisions | 4.16 (4) | 4.35 (4) | 0.19 | 0.284 |

* Significant at p < 0.05 (Mann-Whitney U statistical test)

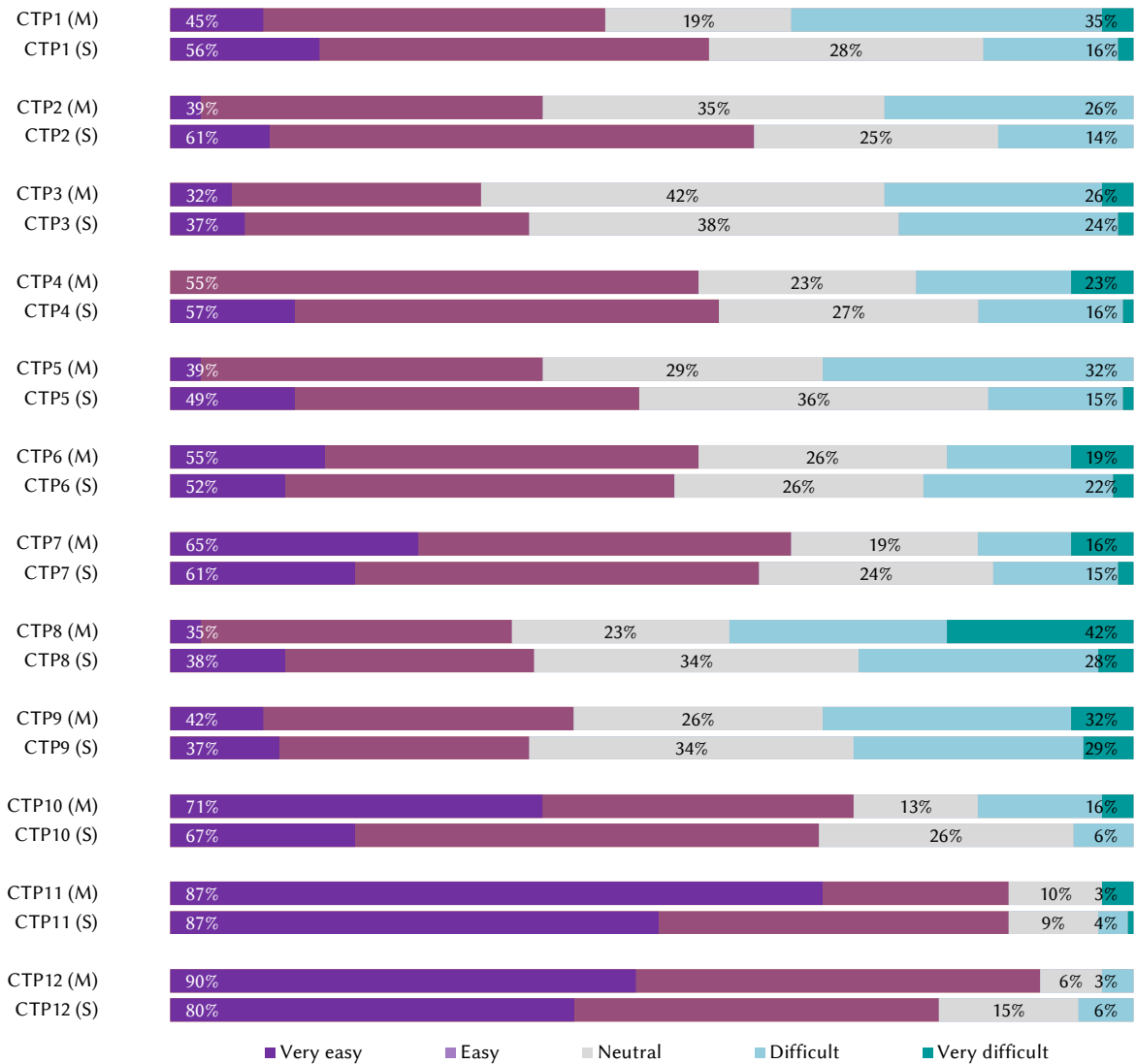

Figure 6. Secondary school girls' (n=193) and mentors' (n=31) perceptions of CT practices (M=Mentor and S= Student)

> ***Finding 1***. Students and mentors do not differ significantly in perceiving the difficulty level of applying computational thinking practices.
>
> ***Finding 2***. The aggregated students' and mentors' opinions indicate that "debugging" (i.e., measured by the statements "finding a solution to fix the identified errors in code" and "identifying errors in code") is the most difficult computational thinking practice to apply, followed by "abstraction" (i.e., measured by the statement "leaving out the irrelevant detail/information in the description of an idea").
>
> ***Finding 3***. From the participants' perspective, collaborative practices of computational thinking, including "reaching a consensus in group decisions", "working collaboratively with team members", and "giving feedback to teammates and making suggestions to improve idea/code" are the easiest practices to apply.

## 4.3 Is there a relationship between secondary school girls' prior coding experience and their perceptions of the difficulty of CT practices? (RQ2)

As previously noted, we collected the level of students' coding experience in the pre-workshop survey. To understand the effect of students' coding experience in their perceptions of CT practices, we first divided students' data into three independent groups. The first group included 69 students who had no coding experience before the OzGirlsCT workshops (See Table 4). The second group consisted of 96 students who had less than 6-month coding experience (i.e., relatively low experienced group). The third one included the students who had relatively moderate experience in coding, indicating more than 6-month experience (i.e., 28 students). Then, the Kruskal-Wallis one-way ANOVA tests [89] were conducted on 3 independent variables (*no experience* group, *relatively low experience* group, and *relatively moderate experience* group) and 12 dependent variables (CTP1...CTP12). The results of the Kruskal-Wallis one-way ANOVA tests are summarized in Table 4.

We observe that the level of coding experience significantly affects how students perceived CTP8 and CTP9. The students who indicated higher coding experience had lower difficulty in "*identifying errors in code*" (CTP8, $\chi^2(2)$ =9.987, p-value=0.007) and "*finding a solution to fix the identified errors in code*" (CTP9, $\chi^2(2)$ =7.790, p-value=0.020). The results of the pairwise post hoc tests (i.e., pairwise comparisons) are shown in Table 5 and Table 6, in which each row tests the null hypothesis that the distributions of each pair group are equal. Concerning the perception of CTP8, we observe that a statistically significant difference exists between the group who had no experience in coding and those who had moderate coding experience (adjusted *p-value= 0.0012*, adjusted using the Bonferroni correction). As shown in Table 5, there are no differences between the *no experience* group and *relatively low experience* one (adjusted *p-value=0.055*) or *relatively low experience* group and *relatively moderate experience one* (adjusted *p-value=0.603*). We observe the same pattern for CTP9 (See Table 6). The perception of CTP9 differs significantly from the students without any coding experience to the moderately experienced students in coding (adjusted *p-value=0.041*).

Table 4. Testing the effect of girls' coding experience in their perceptions of CT practices.

| Variables | Level of Coding Experience | | | p-value |
|---|---|---|---|---|
| | Mean (Median) | | | |
| | No (**n=69**) | Low (**n=96**) | Moderate (**n=28**) | |
| **CTP1**: Planning an idea before implementing it with the micro:bit | 3.55 (4) | 3.52 (4) | 3.61 (4) | 0.931 |
| **CTP2**: Breaking down an initial idea into smaller, more manageable steps/parts | 3.51 (4) | 3.60 (4) | 3.61 (4) | 0.680 |
| **CTP3**: Leaving out the irrelevant detail/information in the description of an idea | 3.14 (3) | 3.27 (3) | 3.04 (3) | 0.486 |
| **CTP4**: Developing a general solution that can be applied to other problems in the future | 3.45 (4) | 3.65 (4) | 3.32 (3.5) | 0.198 |
| **CTP5**: Creating a series of ordered steps to implement an idea with the micro:bit | 3.33 (3) | 3.46 (3) | 3.75 (4) | 0.206 |
| **CTP6**: Exploring diverse solutions to an idea, until the ideal solution is achieved | 3.29 (3) | 3.53 (4) | 3.25 (3.5) | 0.228 |
| **CTP7**: Testing code frequently to check if it works | 3.48 (4) | 3.68 (4) | 3.93 (4) | 0.083 |
| **CTP8**: Identifying errors in code | 2.88 (3) | 3.27 (3) | 3.57 (3) | **0.007*** |
| **CTP9**: Finding a solution to fix the identified errors in code | 2.86 (3) | 3.25 (3) | 3.50 (3) | **0.020*** |
| **CTP10**: Giving feedback to teammates and making suggestions to improve idea/code | 3.67 (4) | 3.86 (4) | 3.93 (4) | 0.125 |
| **CTP11**: Working collaboratively with team members | 4.26 (4) | 4.36 (5) | 4.43 (5) | 0.341 |
| **CTP12**: Reaching a consensus in group decisions | 4.00 (4) | 4.25 (4) | 4.25 (4) | 0.103 |

* Significant at p < 0.05 (Kruskal-Wallis one-way ANOVA)

Table 5. Pairwise Comparisons for CTP8 (grouped based on the level of coding experience)

| | Test Statistic | Standard Error | Standardized Test Statistic | p-value | Adjusted p-value |
|---|---|---|---|---|---|
| No Exp. vs. Low Exp. | 20.01 | 8.49 | 2.357 | 0.018 | 0.055 |
| No Exp. vs. Moderate Exp. | 34.786 | 12.053 | 2.886 | 0.004 | **0.012*** |
| Low Exp. vs. Moderate Exp. | -14.775 | 11.553 | -1.279 | 0.201 | 0.603 |

* Significant at p < 0.05

Table 6. Pairwise Comparisons for CTP9 (grouped based on the level of coding experience)

| | Test Statistic | Standard Error | Standardized Test Statistic | p-value | Adjusted p-value |
|---|---|---|---|---|---|
| No Exp. vs. Low Exp. | 18.747 | 8.499 | 2.206 | 0.027 | 0.082 |
| No Exp. vs. Moderate Exp. | 29.727 | 12.065 | 2.464 | 0.014 | **0.041*** |
| Low Exp. vs. Moderate Exp. | -10.98 | 11.565 | -0.949 | 0.342 | 1 |

* Significant at p < 0.05

> **Finding 4**. Having prior experience in coding significantly reduces the difficulty level of "identifying errors in code" and "finding a solution to fix the identified errors in code".
>
> **Finding 5**. We have no evidence to suggest that the level of coding experience affects the perceived difficulty level of any other computational thinking practices.

## 4.4 What challenges do secondary school girls face when collaboratively implementing computational ideas with the micro:bit? (RQ3)

To understand the challenges and barriers that students experienced while developing and implementing computational ideas with the micro:bit, we asked students the following question: "*What issues and challenges did you face when implementing your ideas with the micro:bit?*" We also solicited mentors' perspectives and observations in this regard. We identified six categories of challenges. Figure 7 visualises the identified challenges. This visualisation enables a reader (i.e., researcher or practitioner) to quickly get an overview of the challenges faced by girls when developing and implementing a computational idea with the micro:bit. In this figure, the outer layer influences all inner layers. For example, as described later in this section, some of the challenges related to "Incorporating idea into the micro:bit" or "Code debugging" can be attributed to the limitations of the micro:bit device. Furthermore, Figure 7 shows that there might be dependencies among the challenges (e.g., increase) in a layer. For example, the complexity of code can increase the challenges related to code debugging. Below, we describe each of these six challenges. For brevity, we only include a few quotations from our participants. Table 7 sorts these challenges based on the frequency of their appearance in the data and shows three illustrative quotations corresponding to each of these six challenges. Table 7 also indicates the frequency of sub-challenges (also presented in Figure 7) that appeared in each identified challenge. Note that the summation of the frequency of sub-challenges in a challenge is sometimes less than the total number of that challenge. For example, while the "incorporating idea in the micro:bit" challenge appeared in total 76 times in our data, we only found three reasons (sub-challenges) behind this challenge: "breaking down or narrowing an idea" (n=18), "creating a series of ordered steps to implement an idea" (n=17), and "leaving out irrelevant/unimportant stuff from an idea" (n=8). For the rest of the references (N=33) referring to this challenge, while the participants mentioned this challenge, they did not provide any reasons behind it.

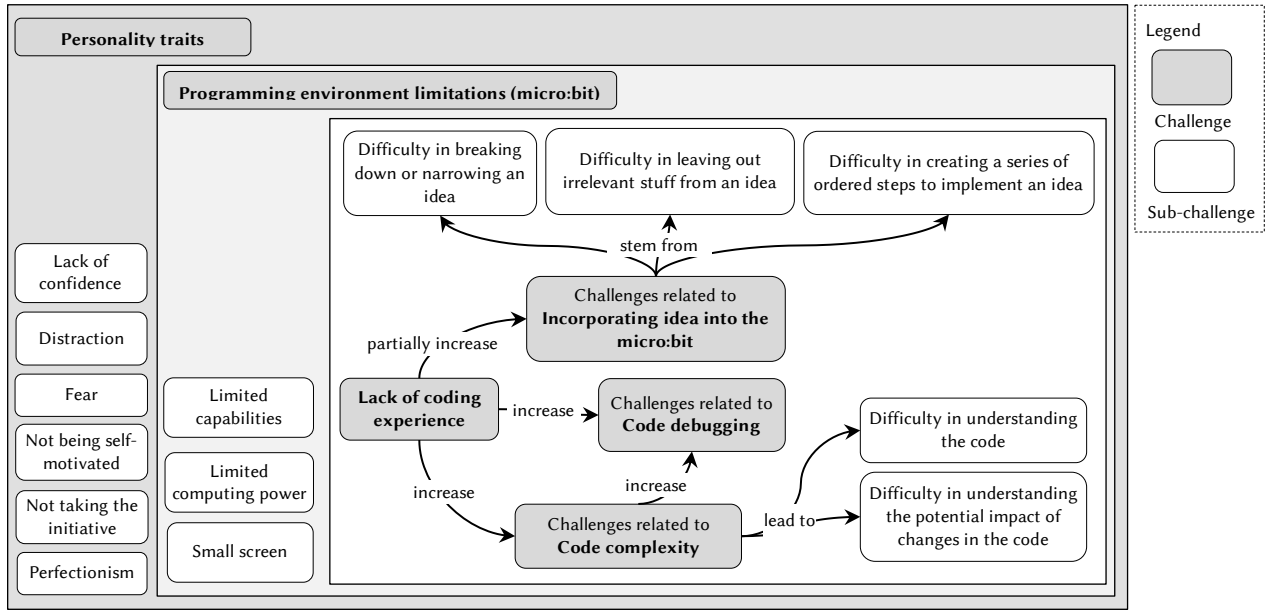

Figure 7. An overview of challenges that that secondary school girls faced when working on computational solutions and the relationship among them.

Table 7. Illustrative quotes for the challenges that secondary school girls faced when working on computational solutions (**N** shows the frequency of each challenge that appeared in total in the data)

| Theme (challenge) | N | Illustrative Quotes |
|---|---|---|
| **Incorporating idea into the micro:bit**<br><br>*Breaking down or narrowing an idea* (**18**)<br>*Creating a series of ordered steps to implement an idea* (**17**)<br>*Leaving out irrelevant/unimportant stuff from an idea* (**8**) | 76 | "*Certain ideas we had could not be properly interpreted on the micro:bit.*" **SS63**<br>"*As many of them [students] had a little coding experience, they often had no idea how to translate their ideas and concepts into smaller, manageable steps to be implemented into the code.*" **MS12**<br>"*Choosing what specific aspects [of the idea] to display on the micro:bit [was a challenge].*" **SS12** |
| **Code debugging** | 41 | "*The code had a few confusing errors that I didn't know how to fix.*" **SS82**<br>"*They struggled with a small piece of code that seemed to malfunction. Being able to test what part of the code was going wrong was a barrier, but that could be rectified with some more practice in debugging code.*" **MS14**<br>"*They often stuck on the errors and got so frustrated when they believe they cannot solve the problem.*" **MS29** |
| **Personality traits**<br><br>*Lack of confidence* (**7**)<br>*Distraction* (**6**)<br>*Fear* (**5**)<br>*Not being self-motivated* (**3**)<br>*Not taking the initiative* (**2**)<br>*Perfectionism* (**2**) | 25 | "*The girls often got side-tracked and overexcited.*" **MO14**<br>"*At the end when the 2 of them were getting easily distracted, 1 of the other girls (the writer) would ask the girls to focus and pull their attention back*". **MO7**<br>"*[Students] waited for me to do something, sat there, [and they were] quiet until I did something.*" **MO2** |
| **Code complexity**<br><br>*Difficulty in understanding the code* (**9**)<br>*Difficulty in understanding the potential impact of changes in the code* (**4**) | 22 | "*[The team members were] not testing small parts and then realising after a big piece of code was written that it didn't work as expected and getting frustrated.*" **MS15**<br>"*Even if I wanted to change things in the code (which required changing in the JavaScript form), it was difficult for me.*" **SS511**<br>"*We faced some problems with getting all of the code to work.*" **SS291** |

| Micro:bit limitations<br><br>Limited capabilities **(14)**<br>Limited computing power **(5)**<br>Small screen **(3)** | 22 | *"It [micro:bit] doesn't have everything that would actually make what we wanted, however, we managed to make something similar to what we wanted to address the issues and challenges, the group consulted with each other and asked our mentors for help when we reached a stump in our thinking."* **SS201**<br>*"We also had some issues with the micro bit as it was not as flexible as we would like it to be in terms of functionality."* **SS224**<br>*"This was hard to implement on the micro:bit device as the device has a limited computing power."* **MO9** |
|---|---|---|
| **Coding experience** | 21 | *"I think it was simply the lack of experience I had with coding (besides the pre-work and basic coding in Year 7) and my unfamiliarity with the functions of the coding blocks that made it quite difficult to implement the ideas with the micro:bit."* **SS161**<br>*"There were still a few variables that I was not known of."* **SS503**<br>*"They have limited education about programming; thus, they only could make the prototype of their product."* **MS26** |

**Incorporating idea into the micro:bit**. The most frequently reported challenge was transferring the planned idea to the micro:bit (e.g., *"It was a bit difficult to convert my worded problem into one which the code could follow."* **SS11**). Our analysis shows that this challenge mainly stems from the fine-grained challenges expressed in the open-ended questions, including the **difficulty** in "breaking down or narrowing an idea", "leaving out irrelevant/unimportant stuff from an idea", and "creating a series of ordered steps to implement an idea". In Section 4.2, the participants rated that it was *moderately* difficult to apply CTP2, CTP3, and CTP5, which are linked to these fine-grained challenges, respectively. We found frequent references in our data sources that students struggled to break down their ideas in a way that could be implemented with the micro:bit. Students also frequently complained about how and where to start coding (e.g., *"I didn't have enough coding knowledge to know where to start, so I struggled with such a challenge."* **SS22**). These challenges seem to be magnified by a lack of coding experience. The simplicity of the micro:bit meant that students had to essentially leave out the unnecessary details and features from their ideas and solutions. **MO20** pointed out:

> *"The team initially wanted to jam-pack their idea with many features. However, I had to bring it down for them to decide which features they thought were the most important and to focus on a few key features that would really sell the idea."* **MO20**

**Code debugging.** As we discussed in Section 4.2, our respondents rated the testing- and debugging-related practices (i.e., CTP7, CTP8, and CTP9) as the most difficult CT practices to apply. Through the open-ended questions in the surveys and observation study, many respondents attempted to provide clarifications and reasons for their answers to CTP7, CTP8, and CTP9 (e.g., *"It was hard to fix the errors since we didn't actually know what the error in the code was"* **SS124**). One mentor, **MS14**, pointed out these challenges could be mitigated by gaining more experience in code debugging. According to **MS8**, **MS15**, and **M15**, the inability to quickly resolve the issues with the code seems to generate frustration and demotivate the students. The students tried *not* to make even simple mistakes in the code because they felt they could not solve the issues in the code. One student elaborated on:

> *"When we found flaws in our product we were creating, particularly ones that we took a long time (or didn't manage at all) to find solutions for, it kind of demotivated me for a little while."* **SS163**

**Personality traits.** Besides the technical problems, the respondents reported a group of personality traits as challenges. This confirms the findings of [1, 2] that the learning outcomes of a learner in CT are related to the learner's characteristics (i.e., non-cognitive side of CT). Mentors observed that some students did not have complete confidence in themselves. This resulted in students who *could not* or *did not want* to explore the problem (i.e., ideation) and solution (i.e., coding) spaces properly. One mentor explained how a lack of confidence led to self-filtering in the ideation phase within a team:

> *"Students were quite unsure of themselves and tended to dismiss their ideas. While no idea was initially written down then discarded, I suspect that their ideation phase had a lot of self-filtering going on as students tended to stare intently at their post-it notes. There was also quite a lot of self-doubt, especially when working on their prototype coding."* **MO13**

This lack of confidence can be exaggerated by fear of suggesting something wrong or being wrong (e.g., *"[My students were] too scared to explore functions and do things on their own"* **MS2**). Distraction was another issue related to the personality traits, as some mentors described that it was challenging for students to remain focused. Fewer mentors also specified that students were not self-motivated and did not show initiative or they were exceptionally perfectionist (e.g., *"Some of the members were perfectionists, and I had to stop them from completely erasing their idea when they were struggling to find a solution"* **MO22**).

**Code complexity.** The respondents frequently shared experiencing issues due to the complexity of code on understanding the code, as mentioned by **SS131**: "*We found it hard to make our solution [code] clear and uncomplicated*". **SS311** shared that it was challenging to "*learn and remember what codes did what and in what order they had to go in*". Other students struggled to figure out the potential impact of changes in the code, as mentioned by **SS51**: *"[It was difficult to] learn what code did what and learning if we added a code it affected everything"*.

**Micro:bit limitations.** The participants reported challenges regarding the micro:bit device frequently. Among these, the limited capabilities of the micro:bit was the prominent one, especially from the students' perspective. Here is one of the examples indicating students believed that the limited capabilities of the micro:bit did not allow them to completely implement their ideas:

> *"When implementing our ideas with the micro:bit, we struggled with its capabilities and spent a bit of time researching how to solve these limits. In the end, we made a very simplified version [of our idea], accepting that the micro:bit could not handle the final product."* **SS301**

Some respondents described the computing challenges related to the micro:bit and reported that the micro:bit device had limited computing power to execute the complex code (e.g., "*The micro:bit does not have the capability to execute code with the level of complexity that my code possessed*" **SS302**). Having a small screen and a limited number of functions were also reported as issues related to the micro:bit.

**Coding experience.** In some cases, the (technical) challenges that emerged from our data related to the students' experience and expertise in coding. Having adopted the experiential learning approach, some respondents reported that it was challenging to implement the ideas with the micro:bit and ensure the code worked properly with the limited coding experience and knowledge that students had. This can be vividly exemplified by the following quote:

> *"I think it was simply the lack of experience I had with coding (besides the pre-work and basic coding in Year 7) and my unfamiliarity with the functions of the coding blocks that made it quite difficult to implement the ideas with the micro:bit."* **SS161**

> ***Finding 6***. *According to our participants, incorporating an idea into the micro:bit device as a hybrid block/text programming environment is the most challenging task while developing and implementing a computational idea.*
>
> ***Finding 7***. *Participants provide evidence that the challenges and barriers faced when developing and implementing computational ideas can also be attributed to personality traits. This qualitative result is consistent with the existing literature [1, 2], which quantitatively reveals the existence of a non-cognitive side of computational thinking.*

## 4.5 What practices do secondary school girls employ to overcome these challenges? (RQ4)

The practices and techniques that students employed to overcome the challenges experienced during the ideation and coding were collected through a compulsory open-ended question in each of our data sources: the student survey, the mentor survey, and the observation protocol. Our qualitative analysis described in Section 3.3.2 revealed that students used six main practices during the ideation and coding using the micro:bit device. Figure 8 shows these six practices and indicates which challenges presented in Section 4.4 are (partially) addressed by these practices. We further visualise the relationship among these practices in Figure 9. In Figure 9, the outer layer influences all inner layers. This section defines, elaborates, and provides examples of each practice. We describe the six identified practices with a few quotations from our participants. For brevity, we only include a few quotations from our participants. Table 8 sorts these practices based on the frequency of their appearance in the data and shows three illustrative quotations corresponding to each of these six practices.

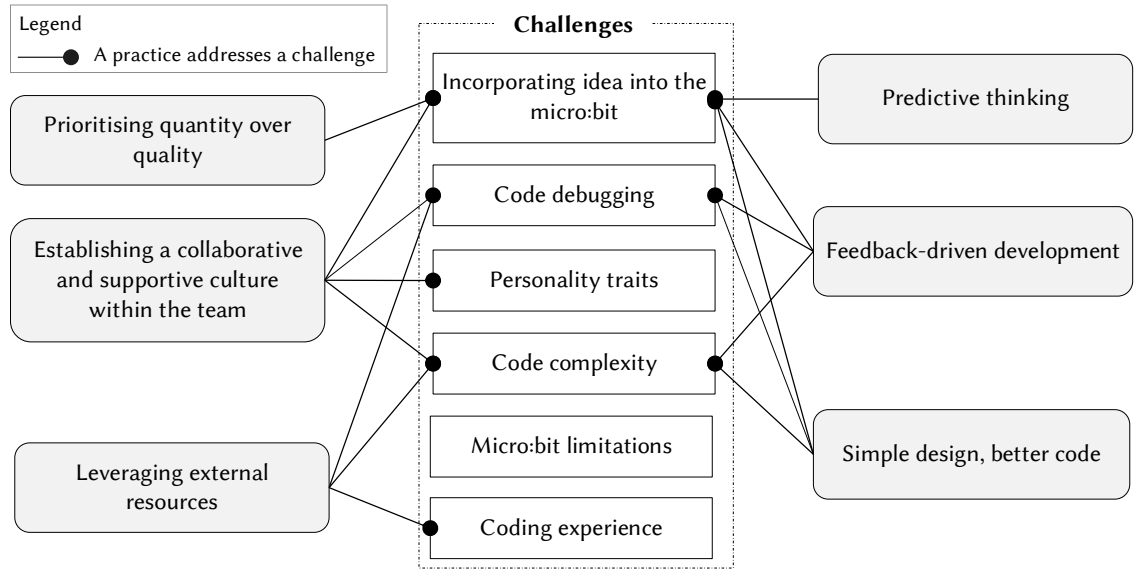

Figure 8. The relationship between the identified challenges and practices

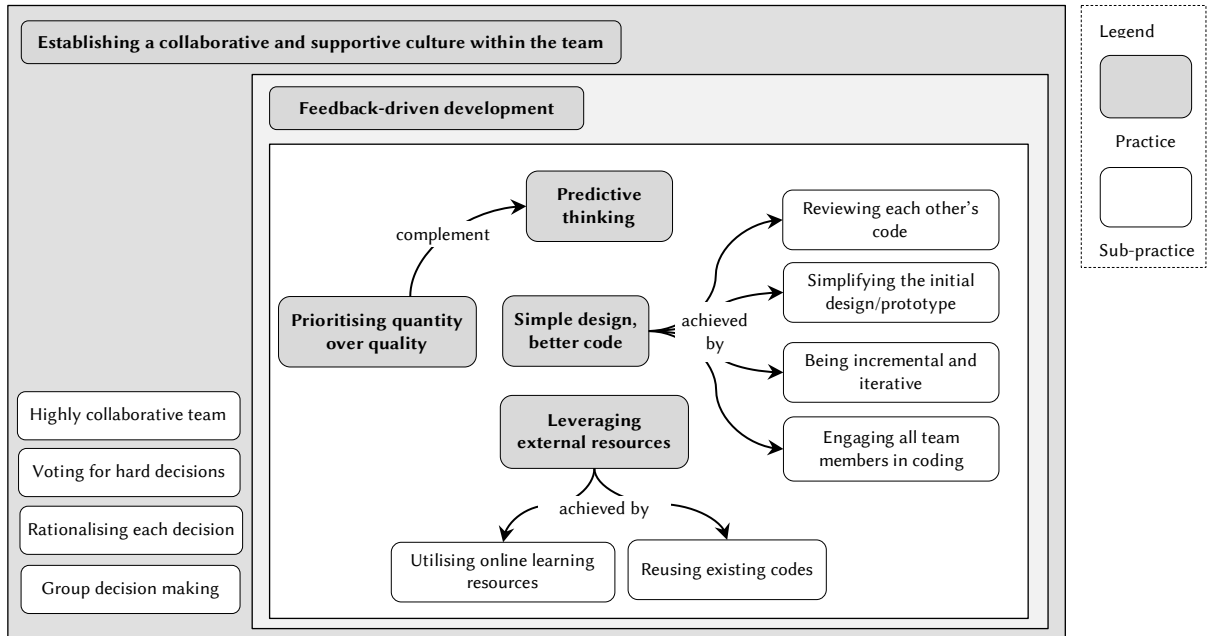

Figure 9. An overview of practices employed by secondary school girls and the relationship among them.

Table 8. Illustrative quotes for the practices employed by secondary school girls (**N** shows the frequency of each practice that appeared in total in the data)

| Theme (practice) | N | Illustrative Quotes |
|---|---|---|
| **Feedback-driven development** | 72 | *"We asked [our mentor] to help us figure out what was not working in the code and she really helped us link everything together."* **SS321**<br>*"We discussed in a team the possible ways we could overcome the issues so that we all had different ideas. We also talked to our tribe about these issues to get other people's opinions."* **SS212**<br>*"The team talked with one another and provided comments on each other's ideas."* **MO23** |
| **Establishing a collaborative and supportive culture within the team**<br>*Highly collaborative team* **(45)**<br>*Group decision making* **(8)**<br>*Voting for hard decisions* **(5)**<br>*Rationalising each decision* **(5)** | 63 | *"With the obstacles [that] we had to overcome along the way, we're a strong team with a great work ethic and I could see this project going beyond a prototype."* **SS81**<br>*"The team fixes issues in a collaborative manner whereby each member provides constant input on possible solutions, which is then explored by the entire team."* **MO8**<br>*"They thought critically for some justifications for their cost problem and were able to find a way for their justification. Each decision they made was always discussed among each other."* **MO28** |
| **Simple design, better code**<br>*Simplifying the initial design/prototype* **(25)**<br>*Being incremental and iterative* **(11)**<br>*Engaging all team members in coding* **(4)**<br>*Reviewing each other's code* **(3)** | 43 | *"After solving the problems, they sought to further improve the product and patch up any inconsistencies. There was a lot of trial-and-error with the coding. There was evidence of iterative approaches to the problems."* **MO17**<br>*"[Our solution was to] work together and checking each other's' code."* **SS522**<br>*"They tended to keep it pretty simple and only showed a part of the end idea (app)."* **MS11** |
| **Predictive thinking** | 30 | *"They [team] brainstormed on what they all thought should be included [in the final design] on the whiteboard, so they could visualise and discuss much clearer. Based on this, they were able to draw up what they thought they wanted the final design to look like."* **MO5**<br>*"[They] kept brainstorming and spending time developing the ideas and making sure that it was as strong as it could be."* **MO46**<br>*"We did not find many challenges as we chose an idea that would work with the micro:bit."* **SS524** |
| **Leveraging external resources**<br>*Utilising online learning resources* **(22)**<br>*Reusing existing codes* **(6)** | 28 | *"I also decided to research some tutorials or general coding techniques to use and asked our mentor for some advice when needed."* **SS161**<br>*"[We were] using the website / pre-workshop guidelines to find connections."* **SS264**<br>*"[They] looked up online articles and copied code to see if it worked."* **MO27** |
| **Prioritising quantity over quality** | 17 | *"To solve a problem, we talked as a group and tried our best to give as much as ideas and pick the most suitable one for the issue."* **SS151**<br>*"We asked questions and also tried different codes over and over again."* **SS124**<br>*"[The team was] thinking of possible ways to make it as realistic as possible."* **MS16** |

**Feedback-driven development.** Our analysis shows that the most common practice used by students was the feedback-driven development practice (i.e., seeking help and feedback) [99], in which students constantly received feedback by consulting with their team members and mentors. Students mainly used the feedback-driven development practice to enhance the quality of their code and ideas. **SS201** commented on the importance of the received feedback during the ideation phase, as shown by the following:

> *"To address the issues and challenges, the group consulted with each other and asked our mentors for help when we reached a stump in our thinking."* **SS201**

Whilst students mainly provided feedback to and received feedback from their teammates and mentors, some students attempted to obtain authentic and diverse feedback from the members of other teams as well. Our observation shows that as the workshop day progressed, students were more willing to seek feedback and ask more questions. According to **MO31**, this was an effective practice as students became conscious of seeking help instead of backing out.

**Establishing a collaborative and supportive culture within the team.** Several participants discussed the effects of working as a team to address the challenges and issues. We found that a collaborative and supportive team implies that the team explores the issues collaboratively through scenarios and subsequently resolves the issues in that context whereby each team member evenly provides inputs on possible ideas and solutions. A mentor described the characteristics of the collaborative and supportive team as follows:

> *"The issue that they encountered were discussed evenly across the members with each providing their own take on the issue that was considered equally by each member."* **MO10**

The quantitative analysis of three single-choice questions from the observation reports confirms that the majority of the teams were successful in establishing a collaborative and supportive culture. As shown in Figure 10, mentors observed that the students of 39 teams out 52 teams *evenly* contributed in their respective teams, whilst other teams either were relying on one "leader" to carry out all the tasks (1 team) or had a dominant member who guided the work process and other students contributed *unevenly* (12 teams). Figure 11 shows the communication style within the teams in two time-slots (i.e., ideation time and coding time). It shows that the level of the communication within the teams was reasonably high, as over 78.8% of the teams adopted the style where all the students communicated *freely* or *well*, in which critical evaluations, objections, critiques, and opinions were freely exchanged within the teams. According to Figure 11, the level of communication increased between the ideation time and the coding time.

We found a number of statements from students and mentors about the effects of group decision-making in addressing the encountered issues. Voting was deemed by a few participants as an effective practice to deal with hard decisions (e.g., *"[The team] narrowed ideas down via voting"* **MO1**). Some teams tried to rationalise each decision and incorporate all team members' opinions in the decisions made.

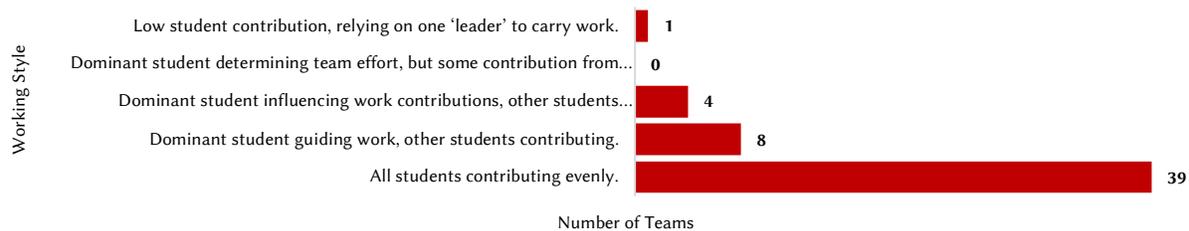

Figure 10. Teams' working style

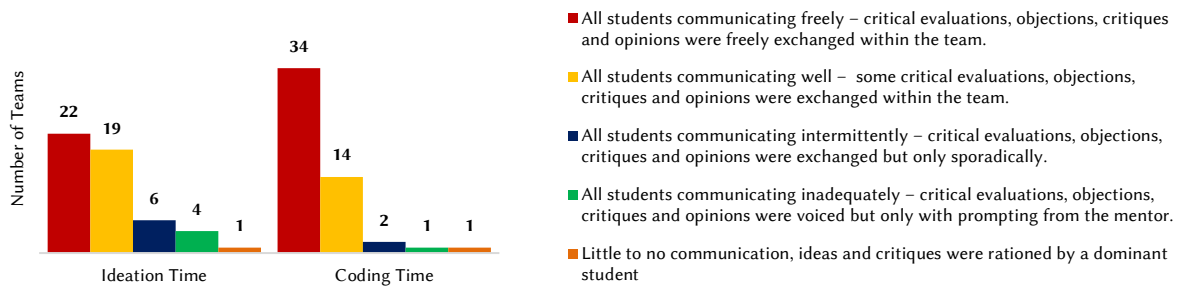

Figure 11. Teams' communication styles

**Simple design, better code.** Throughout the workshop day, students learned that most of the difficulties that they faced with "incorporating ideas into the micro:bit" and "testing and debugging the code" (See Figure 8) could be addressed by simplifying the initial design and incrementally adding new functionality into the code. Our participants reported that it was originally challenging to incorporate the micro:bit in the original design (prototype); however, after simplifying the design, students were able to use the micro:bit effectively. One student put it as:

> *"We asked ourselves if certain things were necessary for the prototype or if we should just take it to the basics to show our idea and then in further development, we can add more detail."* **SS91**

Having a simple design and being incremental and iterative enabled students to write code that was more manageable and testable (e.g., *"Instead of trying to replicate the whole app on the micro:bit, [the team] showing just a few functionalities on the micro:bit, [which] made the code a lot more manageable"* **MO24**). According to **MO17**, this method of problem-solving was employed by her team to track down the root cause of errors. The team was able to resolve the logical errors in their code by breaking the code down into smaller components and testing each of them individually. We found that a few teams could fix errors in the code by engaging all team members in coding and reviewing each other's code. This can be exemplified by the following quote:

> *"We made sure that everyone in our team got to code so that we could each figure out a way to fix."* **SS473**

**Predictive thinking.** The challenges faced by students when incorporating their ideas (prototypes) into the micro:bit device taught them that they need to spend more time reflecting on their ideas and predicting the potential outcomes of their ideas before actually implementing the ideas with the micro:bit [100] (e.g., *"It [our technique] was experimenting with the different blocks and predicting which ones would be applicable to the design that we had envisioned"* **SS161**). For example, **SS251**, **SS372**, and **SS524** shared that their respective teams did not face any significant challenges during implementation and the reason behind was that they first planned all different ideas concretely and then chose an idea that could be implemented with the micro:bit.

**Leveraging external resources.** Another notable technique used by students was to browse and leverage external resources when they were unable to solve code-related issues. This practice can also be related to a broader practice in CT called "remixing and reusing" [12]. The most commonly used resources were micro:bit-related tutorials and websites. For example, **MS20** described that a YouTube tutorial found by her team helped them create an object sensor with the micro:bit. Fewer students leveraged the existing code to add new functionality to their computational solutions. As an example, we have:

> "Another team member was trying to code by copying the sample code but did not understand how the code works. Once she understood how the code works, she got the code done very quickly." **MO3**

**Prioritising quantity over quality.** We discussed that developing a robust, concrete idea before implementing it with the micro:bit was perceived helpful by some students and mentors to reduce the challenges experienced when transferring ideas to the micro:bit. However, this could lead students to fixate on a few ideas or solutions [101, 102]. As a complement practice to "predictive thinking", we observed that students tried to explore the design space of their defined open-ended problems as much as possible and to iteratively diverge and converge the computational ideas and solutions around the problems without deeply thinking about quality, eventually selecting a best-fit one. Indeed, this is often referred by both students and mentors as an effective practice to mainly meet the ideation phase's challenges. One mentor pointed out:

> "The ideation issues were solved by the other team members and myself encouraging the students to list down problems or solutions regardless of being solvable or easy to implement." **MO3**

> **Finding 8.** We discovered six main practices and techniques that novice girl programmers found to be the most effective practices to deal with the challenges in collaborative ideation and coding. Whilst, we confirm that some of these practices (i.e., "predictive thinking" and "leveraging external resources") are already revealed/discussed by the literature [3] as a computational thinking practice, four of these practices including "feedback-driven development", "establishing a collaborative and supportive culture within the team", "simple design, better code", and "prioritising quantity over quality" have not been frequently and rigorously examined in the computational thinking literature.

## 5 Discussion

CT educational programs are increasingly being designed to increase women's interest, engagement, and participation in computing and technology fields. Challenging stereotypes and structural and cultural factors that negatively shape women's enthusiasm and abilities around computing is an important step in dealing with the representation issues in STEM [35, 103]. With this in mind, we designed and conducted a girls-only CT program (i.e., the OzGirlsCT program) with one of the aims being to explore the effect of the OzGirlsCT program on the secondary school girls' CT practices when they do collaborative programming with the micro:bit. The feedback received from the participating girls shows that the OzGirlsCT program was an inspiring, engaging, and impactful education program. For example, one student said, "*I was surprised by how much I positively benefited from this experience and have learned a lot more than I anticipated. I feel inspired*". Another student pointed out, "*The program has really kick-started my interest in design tech. I have always found it fascinating, but now I know coding and technology can really be a possible career in my future*". The literature shows that girls-only programs can be beneficial to girls- increasing their performance and boosting their self-efficacy and confidence when compared to co-educational programs [32, 33]. Focusing on girls-only computing programs can lead to improved diversity outcomes in the workforce, make labour markets more competitive, and ensure girls have access to the opportunities available in STEM and Entrepreneurship [33, 104, 105]. In the following sections, we first draw some important implications of the findings of this study and the design of the OzGirlsCT program in the context of the literature on women in computing for research and practice. Then, we discuss the limitations of our study.

### 5.1 Implications

**Our study confirmed that debugging is a complex and difficult CT practice**. We quantitively explored the difficulty level of a set of CT practices by gathering the perceptions of 193 secondary school girls and 31 mentors. Our data shows that code debugging measured with the statements "*identifying errors in code*" and "*finding a solution to fix the identified* *errors in code*" was the most difficult CT practice for the secondary school girls to apply (**Finding 2**). The following reasons can be attributed to why debugging was the most difficult CT practice for the secondary school girls:

*(1) Learning and applying debugging need an adequate amount of time.* Many researchers have investigated the role of gender in developing and acquiring CT skills, and the findings in this regard are mixed. Several studies (e.g., [30, 106]) found no significant relationship between CT skills (including debugging) and gender. Regardless of gender and age, it appears that debugging is a complicated CT practice for learners (e.g., [13, 14]), needing a systematic approach and a substantial amount of time to understand the process and develop the necessary skills [3, 23, 100]. Even for experienced software practitioners, debugging is a complex and time-consuming task [107, 108]. This is supported by the outcome of RQ2 (**Finding 4**), which has statistically found that, among all CT practices, having prior knowledge and experience in coding can significantly reduce the difficulty level of debugging. Similarly, Wilcox and Lionelle [61] found that girls gained more benefits than boys from prior coding/computing experience in the introductory computer science course. While the literature findings are mixed about the difficulty level of CT practices for boys and girls, there is stronger support for the significant amount of training time needed to adequately develop and apply some CT skills [25, 109].

*(2) Education programs should be supportive of debugging.* The organisation and activities of the OzGirlsCT program as an education program may have led to the girls experienced more difficulty in debugging. First, as we followed the problem-based learning approach in the OzGirlsCT program, we did not provide the girls with formal debugging training (See Section 3.1.2). Further to this, the OzGirlsCT program was a one-day program and covered several topics and activities (e.g., brainstorming problems, presenting business pitches). Hence, the girls had limited time to work on their prototypes, identify possible errors in their prototypes (**CTP8**), and find a solution to fix the identified errors (**CTP9**). Finally, as the girls had the freedom to choose and work on their projects, many initially chose projects which were complex. Hence, the complexity of the projects may also have increased the difficulty level of debugging for the girls. Given the nature of the OzGirlsCT program and the minimal level of the students' coding experience, the satisfactory development of a complex CT practice like debugging was going to be a challenge.

*(3) Programming learning tools need to provide active help during the debugging process.* If we accept that debugging is a complex CT practice, programming learning tools and environments are expected to provide scaffolding and mechanisms to guide learners through this complexity [110, 111]. However, it seems that most environments (including the micro:bit device which our program used) do not provide guidance, visual clues, and features by which (young) learners can understand the nature of an error, the reason behind it, and the path to resolution [112]. As described in Section 4.4, the girls' challenges when working on their prototypes can also be attributed to the micro:bit limitations. The girls frequently indicated that the micro:bit device did not have some of the capabilities and features to support them in incorporating, implementing, and debugging their ideas. In [14], a survey of 98 teachers who taught programming in code clubs with different languages, such as Scratch, Python, micro:bit, Java, found that debugging was the most commonly reported learning barrier for boys and girls. The study by Wohl et al. [23] observed that girls and boys could better understand the concept of debugging with unplugged and Cubelets sessions than when they used Scratch (a similar learning environment to the micro:bit device). It is because the physicality of unplugged and Cubelets sessions helped students understand why an error happened, while the errors produced in Scratch only showed that the program did not work.

All this indicates that debugging is an especially difficult CT practice and may require specialized and customised educational materials and tools for girls [108, 113]. We propose that learning tools and environments and education programs need to empower girls in facing debugging challenges and guide them through the debugging process. According to Zeller [107], the debugging process consists of seven steps: (1) "track the problem"; (2) "reproduce the problem"; (3) "automate and simplify the test case"; (5) "find possible infection origins"; (6) "focus on the most likely origins"; (6) "isolate the infection chain"; and (7) "correct the defect". In this study, we only focused on two steps of debugging (i.e., locating and correcting faults). Hence, there is a need for a deep investigation to understand how girls perceive each step of the debugging process [113] and how learning tools and environments can support girls through each step. We assert that educational researchers and practitioners should focus on designing CT learning tools and programs that are an ongoing element of the curriculum, much like mathematics, and are introduced early in the education system, as our findings show that experience helps girls deal with complex CT practices such as debugging.

**Our study highlighted the ease with which secondary school girls applied the collaborative aspects of CT.** Collaboration is a key competence in the 21st century [114]. We measured quantitatively the secondary school girls' attitude toward collaborative problem-solving with three statements: "*giving feedback to teammates and making suggestions to improve idea/code*", "*working collaboratively with team members*", and "*reaching a consensus in group decisions*". We have learned that the collaborative practices of CT were the easiest practices for the secondary school girls to apply (**Finding 3**). Studies show that gender may influence the attitude toward and the outcome of collaborative problem-solving [115, 116]. Ardito et al. [31] found that sixth-grade girls concentrate more on group dynamics and metacognitive process when developing and applying CT skills, while boys are more engaged with the operational aspects of building and coding. They observed that girls in sixth grade scored highest in "teamwork/leadership/effective communication" (analogue to collaboration). Our qualitative analysis also confirms **Finding 3** and the observation of [31]. We found having a collaborative and supportive team and seeking feedback from peers and mentors were deemed by the secondary school girls as effective practices in addressing some technical challenges. These two practices also positively impacted other practices employed by the girls in dealing with technical challenges (See Figure 8 and Figure 9). Jun and colleagues in [117] did not take into account gender in CT skills development. They found that among four steps (i.e., design, personalisation, collaboration, and reflection) of design-based learning (DBL), reflection was the most difficult step for elementary school students, followed by collaboration. It was also revealed that in the traditional, direct method, collaboration was the most difficult step. Interestingly, Jun et al. [117] revealed that collaboration was the most popular step among elementary school students in DBL methodology.

Our findings are aligned with the studies (e.g., [32, 118-120]) showing the benefits of incorporating collaboration in CT learning environments, which can help girls deal with complex CT practices and advanced programming tasks. McDowell et al. [120] showed that collaboration could increase girls' persistence in the debugging process. In another study, Buffum et al. [119] and Liebenberg et al. [118] indicated that collaborative learning helped increase girls' enjoyment in CT. The study of Buffum et al. also showed that when girls were paired with a more experienced boy, they were able to solve challenging programming problems as much as the boys. Given the positive influences of collaboration on girls in developing and applying CT skills and our findings regarding the secondary school girls' attitudes toward collaborative problem-solving, we assert that future CT and computer science learning programs for girls should emphasise and encourage collaboration in the program design.

The friendship level of the girls in the teams may have also contributed to the results obtained for the collaborative practices of CT. As discussed in Section 3.1.1.1, while the friendship level of the team members in the teams varied widely, some teams had members who were close friends. Hence, it was not difficult for such teams to work collaboratively on their computational solutions and reach a common goal under pressure conditions. We suggest that the results of the collaborative practices of CT may have been different if the students did not know each other. This necessitates a replication study to investigate how girls who do not know each other perceive the collaborative practices of CT. Finally, further research should focus on possible differences in girl-only, boy-only, and mixed-gender teams in collaborative ideation and programming and in perceiving the difficulty level of the collaborative practices of CT.

**Our study found that focusing on collaborative, creative problem-solving for real-world problems was a strong motivator for our student teams**. Many introductory CT programs focus on programming for pre-defined problems or canned exercises, which may prevent young girls from understanding how computing and technology can help address realistic problems in society [35]. We designed our program with the view of providing opportunities that allowed secondary school girls to collaboratively identify and work on complex problems that are important for them and their society (i.e., real word problems). Working collaboratively on computational projects in personal and social contexts is considered a powerful approach to engage underrepresented populations (e.g., young girls) in computing fields [37, 42, 43, 121]. In the context of CT, this trend is sometimes called "computational action" [43]. Given this opportunity in our program, the secondary school girls developed creative computational solutions to a wide range of personally and socially relevant, realistic problems such as "*how to improve the safety of cyclists*", "*how to reduce food wastage*", "*how to help beginners grow sustainable produce in their own home or garden*", and "*how to help pet owners to take care of pets while they are away*". This process included creative thinking, risk-taking, collaborative problem-solving, and hands-on experiences. The teams collaboratively developed their computational ideas and collected requirements to formulate, implement, and evaluate ideas with the micro:bit. This type of activity positively influences girls' interests in pursuing computing [32].

Observations of our program indicate that exposing the secondary school girls to complex and realistic problems initially puts them in a relatively difficult situation, and appropriate, relevant instruction is required. Their level of coding experience, together with the limitations of the micro:bit device, made it difficult for the girls to implement and debug their sometimes complex ideas with the micro:bit. That is why we found many references in our qualitative data relating to the difficulty of incorporating their idea into the micro:bit device (**Finding 6**). In Section 4.4, we argued that this challenge mainly stems from the difficulty in applying three CT practices: decomposition ("breaking down or narrowing an idea"), abstraction ("leaving out irrelevant/unimportant stuff from an idea"), and algorithm ("creating a series of ordered steps to implement an idea"), in which the difficulty level of abstraction was the second highest amongst our participants (**Finding 2**). Similarly, other studies [110, 122] observed that middle school girls faced barriers in understanding and applying abstraction. Interestingly, the difficulty of incorporating the team's idea into the micro:bit device gradually taught our participating girls that they need to simplify their design, leave out unrelated materials in their initial ideas, and only prototype their solutions with the micro:bit. We referred to this practice in Section 4.5 as "simple design, better code", which was deemed an effective practice to address the challenges related to "incorporating ideas into the micro:bit", "code debugging", and "code complexity" faced by the secondary school girls (See Figure 8). We suggest while developing and working on computational ideas for real-world problems is an excellent motivator and should be considered a strength of CT programs, care should be taken to ensure that young girls understand that they need to 'walk before they can run'. They need to prototype their computational ideas before attempting full implementation, as facing complex programming challenges (e.g., debugging a complex computational solution) early on in the process can be demotivating, and significantly hinder progress, stymieing positive thoughts about courses and careers in computing.

## 5.2 Limitations

Our findings in this study could be influenced by our sampling method. This study focused only on girls from 44 secondary schools in the state of Victoria in Australia. Further, most of our participating students were self-reported high-performers with good academic records. Almost half of them also had at least one month of coding experience before the OzGirlsCT program (i.e., See Figure 5). While an important body of literature (e.g., [14, 25, 30, 31]) shows significant similarities between boys and girls in terms of developing and applying CT practices in different programming languages, our findings are exclusive to secondary school girls with good academic records and prior coding experience. Hence, our study cannot claim that the findings and recommendations be easily transferred or applicable to primary school girls, a broader class of secondary school girls, or students of any gender. Recruiting the participants likely had self-selection bias [123], as the girls and mentors who chose to participate were more likely interested in the study topic. Our findings may be different if this study is replicated in different contexts (e.g., replicating the study with the secondary school girls who do not know each other or are forced to participate in the study regardless of interest). Despite these limitations, we strongly believe that the design of our study and the OzGirlsCT program is not restricted to a particular context, which will facilitate the replication and validation of this study and the OzGirlsCT program in different contexts such as male students or students in different countries.

Another possible limitation emerges from the structure and formulation of the questions used in the surveys and observation study [124, 125]. We formulated the questions used in this study based on a review of formal and grey literature and our prior experience in software engineering research. Furthermore, the questions were fine-tuned and validated through several internal discussions among authors and other colleagues. While we are confident that our questions covered important CT practices that can be introduced and evaluated through programming in K-12 settings, we confirm that some CT practices such as modularity and creativity have not been covered.

The analysis and classification of the qualitative data (i.e., the answers to the open-ended questions) were mainly performed by the first author. Whilst Tómasdóttir et al. [126] argue that this type of analysis process can help increase consistency in the findings, this can be another source of threat to the findings of this study. Our approach to moderate this threat was discussing and cross-validating the categories and their respective quotes with other researchers involved in this study, particularly with the second author. We also minimized the subjective judgement in the findings by reporting those findings that have appeared in more than one data source and reported by multiple participants. Despite these efforts, we cannot claim that we could completely exclude the possibility of interpretation and classification errors.

In this study, the mentors played two roles: facilitator and observer. One potential threat here is that the mentors may have become biased by taking on the role as a member of their teams and becoming involved in the problem-solving process of their teams. Our strategy to deal with this threat was to organise a training workshop for the mentors to clearly define their roles and responsibilities during the OzGirlsCT workshops (See Section 3.1.1.2). Furthermore, during the OzGirlsCT workshops, we occasionally reminded the mentors that they should not be directly involved in the problem-solving process of their teams. Despite these mitigating strategies, it is possible that being an observer may have interfered with their mentoring activities and contributed to the team's frustration, as described in Section 4.4, as they were not as readily available to help.

Social desirability bias [127] (i.e., a participant tends to provide socially acceptable answers) may have influenced the students' and mentors' responses. The potential social desirability bias in the students' responses was mitigated to some extent by asking the mentors the same questions as those of the students. To further reduce this bias, we assured the participants that their personal information and answers would not be identified in any potential reports in the future.

# 6 Summary

This study empirically explored how secondary school girls perceive Computational Thinking (CT) practices by conducting a mixed-methods approach consisting of two surveys with secondary school girls (with 193 valid responses each), a survey with 31 valid responses collected from 31 mentors, and 52 mentor observation reports. Our goal was to understand how secondary school girls perceive the difficulty level of 12 CT practices when developing and implementing computational solutions to socially relevant problems with the micro:bit device in a collaborative setting. We were also interested in understanding the challenges they faced in programming with the micro:bit device in this setting and the best practices they developed and applied to address these challenges.

(1) The quantitative analysis shows that "*identifying errors in code*" (28% of the girls and 42% of the mentors rated it *difficult* or *very difficult*) and "*finding a solution to fix the identified errors in code*" (29% of the girls and 32% of the mentors rated it *difficult* or *very difficult*) are the most difficult CT practices to apply. In contrast, the collaborative practices of CT represented by "*reaching a consensus in group decisions*", "*working collaboratively with team members*", and "*giving feedback to teammates and making suggestions to improve idea/code*" are the easiest practices to apply, as in each case, at least 67% of the participants believe that they are (very) easy practices to apply.
(2) The challenges that the girls faced when developing and implementing their computational solutions with the micro:bit device are not *only* technical (i.e., challenges related to "incorporating idea into the micro:bit", "code debugging", and "code complexity"), but also are rooted in "the micro:bit limitations", "personality traits", and "coding experience".
(3) Whilst having prior experience in coding to some extent influences the difficulty level of CT practices perceived by the girls, its most significant effect is on "debugging".
(4) The main practices employed by the girls to overcome the challenges faced are "feedback-driven development", "establishing a collaborative and supportive culture within the team", "simple design, better code", "predictive thinking", "prioritising quantity over quality", and "leveraging external resources".

## Acknowledgements

This work is funded by the Australian Government, Department of Industry, Innovation and Science, Grant No. WISE64905. The authors would like to thank all participants in this study.

32[92] D. W. Nordstokke, B. D. Zumbo, S. L. Cairns, and D. H. Saklofske, "The operating characteristics of the nonparametric Levene test for equal variances with assessment and evaluation data," *Practical Assessment, Research & Evaluation,* vol. 16, 2011.

[93] A. Field, *Discovering statistics using IBM SPSS statistics.* SAGE Publications Inc., 2013.

[94] B. G. Glaser and A. L. Strauss, *Discovery of grounded theory: Strategies for qualitative research.* Routledge, 2017.

[95] R. Hoda and J. Noble, "Becoming Agile: A Grounded Theory of Agile Transitions in Practice," in *IEEE/ACM 39th International Conference on Software Engineering (ICSE)*, Buenos Aires, Argentina 20-28 May 2017 2017, pp. 141-151, doi: 10.1109/ICSE.2017.21.

[96] R. Hoda, "Self-organizing agile teams: A grounded theory," PhD Thesis, Victoria University of Wellington, 2011.

[97] M. Debnath, M. Pandey, N. Chaplot, M. R. Gottimukkula, P. K. Tiwari, and S. N. Gupta, "Role of soft skills in engineering education: students' perceptions and feedback," in *Enhancing Learning and Teaching Through Student Feedback in Engineering*, C. S. Nair, A. Patil, and P. Mertova Eds.: Chandos Publishing, 2012, pp. 61-82.

[98] B. H. Cohen, *Explaining psychological statistics.* John Wiley & Sons, 2008.

[99] M. Beller, "Toward an empirical theory of feedback-driven development," presented at the 40th International Conference on Software Engineering: Companion Proceeedings, Gothenburg, Sweden, 2018.

[100] G. Falloon, "An analysis of young students' thinking when completing basic coding tasks using Scratch Jnr. On the iPad," *Journal of Computer Assisted Learning,* vol. 32, no. 6, pp. 576-593, 2016, doi: 10.1111/jcal.12155.

[101] S. Dow, J. Fortuna, D. Schwartz, B. Altringer, D. Schwartz, and S. Klemmer, "Prototyping dynamics: sharing multiple designs improves exploration, group rapport, and results," presented at the ACM Conference on Human Factors in Computing Systems, Vancouver, BC, Canada, 2011.

[102] S. P. Dow, A. Glassco, J. Kass, M. Schwarz, D. L. Schwartz, and S. R. Klemmer, "Parallel prototyping leads to better design results, more divergence, and increased self-efficacy," *ACM Transactions on Computer-Human Interaction,* vol. 17, no. 4, pp. 1-24, 2010, doi: 10.1145/1879831.1879836.

[103] S.-C. Kong, M. M. Chiu, and M. Lai, "A study of primary school students' interest, collaboration attitude, and programming empowerment in computational thinking education," *Computers & Education,* vol. 127, pp. 178-189, 2018.

[104] I. Miliszewska and E. M. Sztendur, "Interest in ICT studies and careers: Perspectives of secondary school female students from low socioeconomic backgrounds," *Interdisciplinary Journal of Information, Knowledge, and Management,* vol. 5, pp. 237-260, 2010.

[105] J. Teague, "Women in computing: What brings them to it, what keeps them in it?," *ACM SIGCSE Bulletin,* vol. 34, no. 2, pp. 147-158, 2002.

[106] L. Werner, J. Denner, S. Campe, and D. C. Kawamoto, "The fairy performance assessment: measuring computational thinking in middle school," in *Proceedings of the 43rd ACM technical symposium on Computer Science Education*, 2012, pp. 215-220.

[107] A. Zeller, *Why programs fail: a guide to systematic debugging.* Elsevier, 2009.

[108] M. Perscheid, B. Siegmund, M. Taeumel, and R. Hirschfeld, "Studying the advancement in debugging practice of professional software developers," *Software Quality Journal,* vol. 25, no. 1, pp. 83-110, 2017.

[109] M. U. Bers, L. Flannery, E. R. Kazakoff, and A. Sullivan, "Computational thinking and tinkering: Exploration of an early childhood robotics curriculum," *Computers & Education,* vol. 72, pp. 145-157, 2014.

[110] H. Webb and M. B. Rosson, "Using scaffolded examples to teach computational thinking concepts," in *44th ACM technical symposium on Computer science education*, Denver, Colorado, USA, 2013, 2445227: ACM, pp. 95-100, doi: 10.1145/2445196.2445227.

[111] J. Denner, L. Werner, and E. Ortiz, "Computer games created by middle school girls: Can they be used to measure understanding of computer science concepts?," *Computers & Education,* vol. 58, no. 1, pp. 240-249, 2012.

[112] A. Jordan-Douglass, V. Kumar, and P. J. Woods, "Exploring computational thinking through collaborative problem solving and audio puzzles," in *17th ACM Conference on Interaction Design and Children*, Trondheim, Norway, 2018, 3210766: ACM, pp. 513-516, doi: 10.1145/3202185.3210766.

[113] V. Grigoreanu *et al.*, "Gender differences in end-user debugging, revisited: What the miners found," in *Visual Languages and Human-Centric Computing (VL/HCC'06)*, 2006: IEEE, pp. 19-26.
32